\documentclass[aps,prd,twocolumn,10pt
,
tightenlines,
nofootinbib,showpacs]{revtex4-1}
\usepackage{amssymb,latexsym}
\usepackage{amsmath,amsbsy,bbm}
\usepackage{epsfig,bm}
\usepackage{graphicx,comment}
\DeclareMathOperator{\tr}{tr}

\begin{document}
\def\a{{\alpha}}
\def\b{{\beta}}
\def\d{{\delta}}
\def\D{{\Delta}}
\def\X{{\Xi}}
\def\e{{\varepsilon}}
\def\g{{\gamma}}
\def\G{{\Gamma}}
\def\k{{\kappa}}
\def\l{{\lambda}}
\def\L{{\Lambda}}
\def\m{{\mu}}
\def\n{{\nu}}
\def\o{{\omega}}
\def\O{{\Omega}}
\def\S{{\Sigma}}
\def\s{{\sigma}}
\def\th{{\theta}}

\def\ol#1{{\overline{#1}}}

\def\Dslash{D\hskip-0.65em /}
\def\Dtslash{\tilde{D} \hskip-0.65em /}

\def\CPT{{$\chi$PT}}
\def\QCPT{{Q$\chi$PT}}
\def\PQCPT{{PQ$\chi$PT}}
\def\tr{\text{tr}}
\def\str{\text{str}}
\def\diag{\text{diag}}
\def\order{{\mathcal O}}

\def\cF{{\mathcal F}}
\def\cS{{\mathcal S}}
\def\cC{{\mathcal C}}
\def\cB{{\mathcal B}}
\def\cT{{\mathcal T}}
\def\cQ{{\mathcal Q}}
\def\cL{{\mathcal L}}
\def\cO{{\mathcal O}}
\def\cA{{\mathcal A}}
\def\cQ{{\mathcal Q}}
\def\cR{{\mathcal R}}
\def\cH{{\mathcal H}}
\def\cW{{\mathcal W}}
\def\cM{{\mathcal M}}
\def\cD{{\mathcal D}}
\def\cN{{\mathcal N}}
\def\cP{{\mathcal P}}
\def\cK{{\mathcal K}}
\def\Qt{{\tilde{Q}}}
\def\Dt{{\tilde{D}}}
\def\St{{\tilde{\Sigma}}}
\def\cBt{{\tilde{\mathcal{B}}}}
\def\cDt{{\tilde{\mathcal{D}}}}
\def\cTt{{\tilde{\mathcal{T}}}}
\def\cMt{{\tilde{\mathcal{M}}}}
\def\At{{\tilde{A}}}
\def\cNt{{\tilde{\mathcal{N}}}}
\def\cOt{{\tilde{\mathcal{O}}}}
\def\cPt{{\tilde{\mathcal{P}}}}
\def\cI{{\mathcal{I}}}
\def\cJ{{\mathcal{J}}}

\def\eqref#1{{(\ref{#1})}}

\title{Hadronic Parity Violation at Next-to-Leading Order}
\author{B.~C.~Tiburzi}
\email[]{btiburzi@ccny.cuny.edu}
\affiliation{ Department of Physics,
        The City College of New York,  
        New York, NY 10031, USA \\
Graduate School and University Center,
        The City University of New York,
        New York, NY 10016, USA \\
RIKEN BNL Research Center, 
        Brookhaven National Laboratory, 
        Upton, NY 11973, USA}
\date{\today}
\begin{abstract}
The flavor-conserving non-leptonic weak interaction can be studied experimentally through the observation of parity violation in nuclear and few-body systems.  
At hadronic scales, 
matrix elements of parity-violating four-quark operators ultimately give rise to the parity violating couplings between hadrons, 
and such matrix elements can be calculated non-perturbatively using lattice QCD.  
In this work, 
we investigate the running of isovector parity-violating operators from the weak scale down to hadronic scales using the renormalization group. 
We work at next-to-leading order in the QCD coupling, 
and include both neutral-current and charged-current interactions.
At this order, 
results are renormalization scheme dependent, 
and we utilize 't Hooft--Veltman dimensional regularization. 
The evolution of Wilson coefficients at leading and next-to-leading order is compared. 
Next-to-leading order effects are shown to be non-negligible at hadronic scales. 
\end{abstract}
\maketitle

\section{Introduction}%

The tremendous progress made in understanding weak interactions led to the development of the Standard Model of particle physics. 
Many corners of the Standard Model have been tested to very high precision through experiment. 
The hadronic neutral weak interaction, 
by contrast, 
is the least constrained of all Standard Model currents. 
Such an interaction is difficult to probe; 
because, 
even while largely shielded from electromagnetic contributions, 
it is easily overshadowed by dominant strong contributions. 
One must look to processes that violate strong interaction symmetries to isolate the hadronic weak interaction. 
Flavor-changing neutral currents are suppressed in the Standard Model due to the so-called GIM mechanism~\cite{Glashow:1970gm}. 
To see the hadronic weak interaction through flavor-conserving neutral currents, 
one must look to isovector hadronic parity violation.\footnote%
{
The charged-current interaction makes contributions to isovector hadronic parity violation at the weak scale that are doubly Cabbibo suppressed, 
i.e.~$\propto |V_{us}|^2 / |V_{ud}|^2 \sim 0.05$. 
As we work to next-to-leading order in the QCD coupling, 
we include such subdominant contributions in our analysis. 
At hadronic scales, 
we find that coefficients of strangeness containing $| \Delta I | = 1$ operators are modified by the partity violating charged-current interaction at the 
$10$--$20 \%$
level. 
}
Recent experimental advances give reason for a concerted effort to study hadronic parity violation theoretically using QCD.

Experimentally hadronic parity violation has been sought in nuclear reactions. 
In fact, 
the first search for hadronic parity violation in the nucleon-nucleon interaction was published the same year
(1957)
that parity violation was discovered in the weak decay of nuclei~\cite{Tanner:1957zz}. 
Since then, 
hadronic parity violation has been observed in a variety of nuclear reactions, 
for example,
the first conclusive evidence came from radiative neutron capture on ${}^{181} \text{Ta}$%
~\cite{Lobashov:1967aa},
and a substantial 
($\sim 10\%$) 
parity-violating effect was observed in 
${}^{139} \text{La}$ 
by using orders-of-magnitude enhancement produced from a near degeneracy of opposite parity states that mix under the weak interaction%
~\cite{Bowman:1989ci}.
A more thorough summary of these and other experiments appears in the reviews~\cite{Adelberger:1985ik,RamseyMusolf:2006dz}. 
Recently bounds on hadronic parity violation in few-body systems have been obtained by measuring the spin rotation of a neutron traveling through 
${}^4 \text{He}$~\cite{Snow:2011zz}, 
and through the measurement a parity-violating spin asymmetry in the reaction $\vec{n} + p \to d  + \gamma$~\cite{Gericke:2011zz}.

Using the experimental data on hadronic parity violation in nuclear reactions to probe the hadronic neutral-current of the Standard Model is an ambitious undertaking. 
Interpretation of the experiments is usually carried out using the meson-exchange model of 
Desplanques, Donoghue, and Holstein (DDH)~\cite{Desplanques:1979hn}. 
In this model, 
various parity-violating meson-nucleon couplings give rise to the parity-violating nucleon-nucleon interaction.
This interaction then generates hadronic parity violation in nuclei. 
The parity-violating couplings are unknown model parameters
(although estimates and broad natural ranges are provided by DDH). 
Assuming the DDH framework, 
constraints on the couplings can be extracted from experiments. 
While the results agree within the reasonable ranges, 
constraints from different parity-violating experiments are not consistent. 
In this framework, 
it is challenging to determine where discrepancies may lie:
in model assumptions buried in nuclear structure calculations, 
in model assumptions about the parity-violating nucleon-nucleon force, 
in dynamical effects due to the non-perturbative nature of QCD; 
or perhaps, 
although very unlikely,  
in the non-leptonic weak interaction itself.

Recently great strides have been made to remove model assumptions about the parity-violating nucleon-nucleon force~\cite{Zhu:2004vw,Phillips:2008hn}.
For few nucleon systems at very low energies, 
one can use effective field theory techniques to parametrize the most general parity-violating nucleon-nucleon interaction based solely on symmetries of the system.\footnote%
{
The effective theory in question is referred to as the pion-less effective theory, see~\cite{Bedaque:2002mn}.
This name obscures the fact that it is a universal low-energy theory valid for systems with large scattering lengths, 
whether they be few-body nuclear systems without pions, atomic systems, \emph{etc}.. 
} 
In this framework, 
there are five unknown parity-violating couplings~\cite{Girlanda:2008ts} 
that can in principle be determined from few-body experiments, 
see, for example,~\cite{Shin:2009hi,Schindler:2009wd,Griesshammer:2011md,Vanasse:2011nd}. 
There is much anticipation that, 
in the not-too-distant future, 
values for some of these couplings will be constrained due to the remarkable experimental program dedicated to measuring hadronic parity violation in few-body systems. 
It is not clear when complete information about the parity-violating nucleon-nucleon interaction will be known. 
Even if the parity-violating nucleon-nucleon force is fully mapped out, 
the connection of its terms to the weak interaction, 
moreover, 
can only be done through non-perturbative QCD.

In lieu of first principles information about matrix elements of these hadronic-scale, 
parity-violating operators, 
symmetry arguments have been used to analyze sources of parity violation at the hadronic level. 
Most notably, 
there was a model-independent analysis of pion-nucleon couplings using chiral symmetry considerations~\cite{Kaplan:1992vj}.
This analysis systematically enumerated the low-energy parity-violating couplings between pions and nucleons, 
as well as pions, photons and nucleons.
Sizable contributions in the 
$|\Delta I| = 1$ 
sector 
(for example, 
to the isovector 
parity-violating pion-nucleon coupling 
$h_{\pi NN}^{(1)}$) 
were argued to arise from operators involving an 
$s \overline{s}$-pair. 
Using chiral perturbation theory, 
the one-loop chiral corrections to 
$h_{\pi N N}^{(1)}$
were later determined%
~\cite{Zhu:2000fc}.
This particular coupling is believed to be phenomenologically important, 
as it gives rise to the long-range part of the parity-violating nucleon-nucleon force.
Theoreticians have pointed to ways this coupling might possibly be measured through high-statistics experiments%
~\cite{Bedaque:1999dh,Chen:2000mb,Chen:2000hb,Chen:2000km}.
Anticipating lattice calculations of the parity-violating pion-nucleon coupling, 
a chiral perturbation theory analysis was presented using partially quenched chiral perturbation theory~\cite{Beane:2002ca}.%
\footnote{
The extension of electroweak charges of quarks advocated by~\cite{Beane:2002ca}, 
however, 
makes the computation difficult to use in practice, 
see~\cite{Tiburzi:2004mv} for further details. 
}

Lattice QCD computations can determine hadronic parity violation from first principles. 
Quite recently the first lattice QCD study of nuclear parity violation appeared~\cite{Wasem:2011tp}.
In this pioneering work, 
the isovector channel was the focus, 
and a signal for the parity-violating pion-nucleon coupling, 
$h_{\pi N N}^{(1)}$,
was obtained.
As with any lattice computation, 
there are a number of systematic errors which must be controlled to make contact with phenomenology. 
As further refinements to the method are made, 
we can expect to see precision information about parity-violating hadronic couplings coming from lattice QCD computations.%
\footnote{
Lattice computations of hadronic parity violation are currently at quite an early stage, 
and are considerably demanding. 
Decades worth of effort has been expended in an analogous pursuit, 
namely the study of matrix elements of flavor-changing weak interactions using lattice QCD. 
Quite recently significant progress has been made towards complete computations, 
for example, 
in the 
$| \Delta S | = 1$
sector, 
see~\cite{Blum:2011pu,Blum:2011ng}.
We are optimistic that such success will carry over into the study of hadronic parity violation in lattice QCD. 
} 
Complete information about hadronic parity violation in few-nucleon systems may not come solely from experiment or theory. 
A combination of the two will thus be essential to provide input for hadronic parity violation in nuclear many-body calculations.
To this end, 
we study how well the sources of parity violation are known at hadronic scales.
A leading-order QCD analysis to this effect was presented some time ago in~\cite{Dai:1991bx}.
Here we focus on the isovector parity-violating operators, 
and include higher-order effects.

Starting from the weak currents in the Standard Model, 
we determine the isovector parity-violating interaction at weak scales to next-to-leading order accuracy in the strong coupling. 
We include contributions from both neutral-current and charge-current interactions. 
The resulting isovector parity-violating four-quark operators are then evolved down to hadronic scales using the 
QCD renormalization group at next-to-leading order. 
To achieve this, 
we borrow heavily from the two-loop formalism extensively developed in the context of flavor-changing weak decays, 
see~\cite{Buchalla:1995vs} for a thorough review. 
We find that next-to-leading order QCD effects on isovector parity-violating operators are non-negligible at hadronic scales. 
Renormalization of parity-violating operators, 
moreover, 
is a key ingredient that must be understood in order to compute
parity-violating hadronic observables on the lattice. 
Although we carry out this investigation using a perturbative renormalization scheme defined in the continuum, 
our findings are nonetheless useful as they can serve as a gateway to convert to other schemes.

Our presentation has the following organization. 
First in Sec.~\ref{s:OB},
we enumerate our operator basis for isovector parity violation. 
This is carried out in three effective field theories: 
one valid just below the weak scale, 
another at intermediate scales, 
and finally the target 
effective field theory valid at hadronic scales. 
In Sec.~\ref{s:C}, 
we determine the isovector parity-violating operators at the weak scale to
next-to-leading order in the strong coupling. 
Results throughout are renormalization scheme dependent,  
and utilize dimensional regularization with the 't Hooft-Veltman scheme, 
in which 
$\gamma_5$
is non-anticommuting~\cite{'tHooft:1972fi,Akyeampong:1973xi,Breitenlohner:1977hr}. 
Next in Sec.~\ref{s:RUN},  
the Wilson coefficients are run down to hadronic scales using the two-loop 
anomalous dimensions. 
Finally in Sec.~\ref{s:LookOut}, 
we summarize our findings, 
and give an outlook to future work.

\section{Operator Basis} \label{s:OB}%

Parity violation in nuclear and few nucleon systems arises from the non-leptonic weak interaction. 
In the isovector channel, 
the leading component stems from exchange of a 
$Z^0$ 
boson, 
with a sub-leading
$\sim 10 \%$ 
contribution from the exchange of 
$W^\pm$ 
bosons. 
At energy scales below the weak interaction, 
we can replace the weak boson exchanges by an effective theory containing contact interactions.

The parity-violating, four-quark operators in the effective theory undergo renormalization from gluon radiation. 
This radiation is important for evolving the operators from the weak scale down to hadronic scales, 
where their matrix elements can be computed using lattice QCD. 
As is standard, 
we must run the five-flavor theory down to the 
bottom quark mass, 
and match onto a four-flavor theory. 
This theory is then run down to the charm quark mass, 
where it is matched onto a three-flavor theory that is valid at hadronic scales. 
We first enumerate a closed basis of operators for each effective theory.

\subsection{Five-Flavor Effective Theory: $m_b <  \mu <  M_{Z} $} %
\label{ss:5}

At scales above the bottom quark mass, 
we have a five-flavor effective theory. 
The isovector parity-violating Lagrangian can be written in the following form
\begin{equation}
\mathcal{L}_{PV}^{(5)}
=
\frac{G_F}{\sqrt{2}}
\sum_{i = 1}^{8}
C^{(5)}_i(\mu) \mathcal{O}^{(5)}_i (\mu)
.\end{equation}
Notice that the effective Lagrangian
$\mathcal{L}_{PV}^{(5)}$
is independent of the renormalization scale
$\mu$. 
The evolution of the Wilson coefficients
$C_i^{(5)} (\mu)$
is exactly compensated by the 
evolution of the operators
$\mathcal{O}_i^{(5)}(\mu)$. 
There are eight such operators.

A complete basis of isovector parity-violating operators in the five-flavor effective theory can be grouped into three sets. 
The first two sets,
\begin{eqnarray}
\mathcal{O}_1^{(5)}
&=&
( \ol u \gamma_\mu u - \ol d \gamma_\mu d )_{L} 
\sum_q^5 ( \ol q  \gamma^\mu q )_{L}
-
\{ L \leftrightarrow R \}
,\notag \\
\mathcal{O}_2^{(5)}
&=&
( \ol u \gamma_\mu u - \ol d \gamma_\mu d \, ]_{L} 
\sum_q^5
 [ \, \ol q \gamma^\mu  q )_L
-
\{ L \leftrightarrow R \}
,\notag \\
\mathcal{O}_3^{(5)}
&=&
( \ol u \gamma_\mu u - \ol d \gamma_\mu d )_L 
\sum_q^5 ( \ol q \gamma^\mu q )_R
-
\{ L \leftrightarrow R \}
,\notag \\
\mathcal{O}_4^{(5)}
&=&
( \ol u \gamma_\mu u - \ol d \gamma_\mu d \, ]_L  
\sum_q^5
[ \, \ol q \gamma^\mu q )_R
-
\{ L \leftrightarrow R \}
,\label{eq:5}
\end{eqnarray}
and
\begin{eqnarray}
\mathcal{O}_5^{(5)}
&=&
( \ol u \gamma_\mu u - \ol d \gamma_\mu d )_L
(\ol s \gamma^\mu s - \ol c \gamma^\mu c + \ol b \gamma^\mu  b )_L
\notag \\
&& \phantom{space}
-
\{ L \leftrightarrow R \}
,\notag \\
\mathcal{O}_6^{(5)}
&=&
( \ol u \gamma_\mu u - \ol d \gamma_\mu d \, ]_L 
[ \, \ol s \gamma^\mu s - \ol c \gamma^\mu c + \ol b \gamma^\mu b )_L
\notag \\
&& \phantom{space}
 -
\{ L \leftrightarrow R \}
,\label{eq:5A}
\end{eqnarray}
arise from $Z^0$ exchange. 
The subscripts 
$L$ 
and 
$R$ 
represent the chiral components of the fermion bilinears, 
so that 
$(\ol \psi \gamma_\mu \psi)_{L,R} = \ol \psi_{L,R} \gamma_\mu \psi_{L,R}$,
with 
$\psi_{L,R} = \cP_{L,R} \psi$, 
and the chiral projectors are defined in the usual way, 
$\cP_{L,R} = \frac{1}{2} ( 1 \mp \gamma_5)$. 
In each case, 
the 
$(\ol \psi \gamma_\mu \psi )$ 
notation indicates a color-singlet contraction, 
$\sum_a \ol \psi {}^a \gamma_\mu \psi^a$. 
On the other hand, 
the mixed bracket notation denotes the following color contraction, 
\begin{equation}
( \ol \psi \gamma_\mu \psi \, ] \, [ \, \ol \phi \gamma^\mu \phi ) = \sum_{a,b} \ol \psi {}^a \gamma_\mu \psi^b \, \ol \phi {}^b \gamma^\mu \phi^a
.\end{equation} 
The chiral basis we employ is different than that chosen for isovector parity violation in~\cite{Dai:1991bx}.
For contributions arising from 
$Z^0$ exchange, 
the chiral basis enables us to eliminate two unnecessary operators in each effective theory. 
This point will be discussed further in Sec.~\ref{s:C}.

Exchange of 
$W$-bosons 
makes a smaller
$\sim 10 \%$
contribution to isovector hadronic parity violation, 
and was not considered in~\cite{Dai:1991bx}.
This contribution is included here
(see Sec.~\ref{s:PVCC} below), 
and the operators arising from 
$W^{\pm}$ 
exchange are chosen to be
\begin{eqnarray}
\mathcal{O}_7^{(5)}
&=&
( \ol u \gamma_\mu u - \ol d \gamma_\mu d )_L (\ol s \gamma^\mu s - \ol c \gamma^\mu c )_L
-
\{ L \leftrightarrow R \}
,\notag \\
\mathcal{O}_{8}^{(5)}
&=&
( \ol u \gamma_\mu u - \ol d \gamma_\mu d \, ]_L  [ \, \ol s \gamma^\mu s - \ol c \gamma^\mu c  )_L
-
\{ L \leftrightarrow  R \}
.\notag\\
\label{eq:5B}
\end{eqnarray}
Taken together, 
the first two sets of isovector parity violating operators are closed under QCD renormalization. 
Additionally the third set is also closed under QCD renormalization.

\subsection{Four-Flavor Effective Theory: $m_c < \mu < m_b$}  %

Integrating out the bottom quark,
we have a four-flavor effective theory. 
The isovector parity-violating effective Lagrangian is written similarly to that above, 
namely
\begin{equation}
\mathcal{L}_{PV}^{(4)}
=
\frac{G_F}{\sqrt{2}}
\sum_{i = 1}^{6}
C^{(4)}_i(\mu) \mathcal{O}^{(4)}_i (\mu)
.\end{equation}
There are now six operators in the effective theory; 
and, 
we choose the basis 
\begin{eqnarray}
\mathcal{O}_1^{(4)}
&=&
( \ol u \gamma_\mu u - \ol d \gamma_\mu d )_{L} 
\sum_q^4 ( \ol q  \gamma^\mu q )_{L}
-
\{ L \leftrightarrow R \}
,\notag \\
\mathcal{O}_2^{(4)}
&=&
( \ol u \gamma_\mu u - \ol d \gamma_\mu d \, ]_{L} 
\sum_q^4
 [ \, \ol q \gamma^\mu  q )_L
-
\{ L \leftrightarrow R \}
,\notag \\
\mathcal{O}_3^{(4)}
&=&
( \ol u \gamma_\mu u - \ol d \gamma_\mu d )_L 
\sum_q^4 ( \ol q \gamma^\mu q )_R
-
\{ L \leftrightarrow R \}
,\notag \\
\mathcal{O}_4^{(4)}
&=&
( \ol u \gamma_\mu u - \ol d \gamma_\mu d \, ]_L  
\sum_q^4
[ \, \ol q \gamma^\mu q )_R
-
\{ L \leftrightarrow R \}
\label{eq:PVC1}
,\end{eqnarray}
and
\begin{eqnarray}
\mathcal{O}_5^{(4)}
&=&
( \ol u \gamma_\mu u - \ol d \gamma_\mu d )_L (\ol s \gamma^\mu s - \ol c \gamma^\mu c )_L
-
\{ L \leftrightarrow R \},
\notag \\
\mathcal{O}_6^{(4)}
&=&
( \ol u \gamma_\mu u - \ol d \gamma_\mu d \, ]_L  [ \, \ol s \gamma^\mu s - \ol c \gamma^\mu c )_L
- 
\{ L \leftrightarrow R \}
.\notag \\
\label{eq:PVC2}
\end{eqnarray}
At these scales and consequently below, 
the contributions from neutral-current and charged-current interactions become indistinguishable
(up to sub-percent contributions we neglect, see Sec.~\ref{s:PVCC} below). 
Furthermore, 
the operators shown in 
Eq.~\eqref{eq:PVC1}, 
and those operators shown in 
Eq.~\eqref{eq:PVC2} 
each form a closed set under QCD renormalization.

\subsection{Three-Flavor Effective Theory: $\L_{\text{QCD}}  < \mu < m_c$} %

Finally at scales below the charm quark mass, 
we match onto a three-flavor effective theory. 
This is the theory used in lattice QCD simulations, 
where the hadronic scale roughly corresponds to the inverse lattice spacing
$\mu \sim a^{-1}$.
The lattice regularization is, 
of course, 
a different scheme than the one considered here. 
The three-flavor isovector parity-violating effective Lagrangian is written the same way as above
\begin{equation}
\mathcal{L}_{PV}^{(3)}
=
\frac{G_F}{\sqrt{2}}
\sum_{i = 1}^{6}
C^{(3)}_i(\mu) \mathcal{O}^{(3)}_i (\mu)
.\end{equation}
For the six terms of this three-flavor effective theory, 
we choose the operator basis 
\begin{eqnarray}
\mathcal{O}_1^{(3)}
&=&
( \ol u \gamma_\mu u - \ol d \gamma_\mu d )_{L} 
\sum_q^3 ( \ol q  \gamma^\mu q )_{L}
-
\{ L \leftrightarrow R \}
,\notag \\
\mathcal{O}_2^{(3)}
&=&
( \ol u \gamma_\mu u - \ol d \gamma_\mu d \, ]_{L} 
\sum_q^3
 [ \, \ol q \gamma^\mu  q )_L
-
\{ L \leftrightarrow R \}
,\notag \\
\mathcal{O}_3^{(3)}
&=&
( \ol u \gamma_\mu u - \ol d \gamma_\mu d )_L 
\sum_q^3 ( \ol q \gamma^\mu q )_R
-
\{ L \leftrightarrow R \}
,\notag \\
\mathcal{O}_4^{(3)}
&=&
( \ol u \gamma_\mu u - \ol d \gamma_\mu d \, ]_L  
\sum_q^3
[ \, \ol q \gamma^\mu q )_R
-
\{ L \leftrightarrow R \}
\label{eq:3PV}
,\end{eqnarray}
and
\begin{eqnarray}
\mathcal{O}_5^{(3)}
&=&
( \ol u \gamma_\mu u - \ol d \gamma_\mu d )_L (\ol s \gamma^\mu s )_L
-
\{ L \leftrightarrow R \}
,\notag \\
\mathcal{O}_6^{(3)}
&=&
( \ol u \gamma_\mu u - \ol d \gamma_\mu d \, ]_L [ \, \ol s \gamma^\mu s  )_L
- 
\{ L \leftrightarrow R \}
\label{eq:SPV}
.\end{eqnarray}
These sets of operators mix with one another under QCD renormalization. 
Furthermore, 
not all of the operators are independent, 
because a Fierz transformation yields the relation
\begin{equation}
\mathcal{O}_2^{(3)} = \mathcal{O}_1^{(3)} - \mathcal{O}_5^{(3)} + \mathcal{O}_6^{(3)}
\label{eq:FC}
.\end{equation}
We will treat this operator as independent, 
however,
and use the Fierz constraint as a consistency check for the anomalous dimension matrices. 
This is only possible because the 't Hooft--Veltman scheme respects Fierz transformations~\cite{Buras:1992tc}.

\section{Isovector Parity Violation at the Weak Scale} \label{s:C}%

Having enumerated an operator basis for isovector hadronic parity violation, 
we now determine the parity-violating component of the non-leptonic weak interaction
at the scale of weak interactions. 
We consider tree-level exchanges of weak vector-bosons, 
as well as the one-loop QCD radiation required at next-to-leading order.

\subsection{Isovector Parity-Violating Neutral Current}

\subsubsection{Tree Level}

In the Standard Model, 
the hadronic neutral current has the form
\begin{equation}
J_{\mu}^{Z^0}
=
\frac{1}{\cos \theta_W}
\left[
\left(
\ol \psi \, \gamma_\mu T^3 \psi
\right)_L
- \sin^2 \theta_W 
\left(
\ol \psi \, \gamma_\mu Q \, \psi
\right)
\right]
\label{eq:Z}
,\end{equation}
where 
$\psi$
is the isodoublet of quark fields
\begin{equation}
\psi = 
\begin{pmatrix}
U \\ D
\end{pmatrix},
\quad
\text{with}
\quad
U = 
\begin{pmatrix}
u \\ c \\ t
\end{pmatrix}, 
\, \text{and} \, \, 
D 
= 
\begin{pmatrix}
d \\ s \\ b
\end{pmatrix}
.\end{equation} 
The third component of weak-isospin, 
$T^3$, 
is given by
$T^3 = \frac{1}{2} \, \diag ( 1, -1 )$, 
and the electric charge matrix has the form
$Q = \frac{1}{6} + T^3$.

At the scale 
$\mu = M_Z$ 
and below, 
we can integrate out contributions from the top quark to arrive at a five-flavor effective theory.
In this theory, 
furthermore, 
the 
$Z^0$
can be simultaneously integrated out, 
leaving only local four-quark interactions. 
At tree level, 
this procedure results in the 
$| \Delta I |= 1$
parity-violating effective Lagrangian given in~\cite{Dai:1991bx}, 
namely
\begin{eqnarray}
\mathcal{L}^{(5)}_{\mu = M_Z}
&=&
\frac{G_F}{\sqrt{2}}
\Bigg[
\frac{1}{3}
\sin^2 \theta_W
( \ol u u - \ol d d)_A \sum_q^{5} (\ol q q)_V 
\notag \\
&& 
+
\left( \frac{1}{2} - \sin^2 \theta_W \right)
( \ol u u - \ol d d)_A (\ol s s - \ol c c + \ol b b)_V
\notag \\
&& 
+
\left( \frac{1}{2} - \sin^2 \theta_W \right)
( \ol u u - \ol d d)_V (\ol s s - \ol c c + \ol b b)_A
\Bigg]
. \label{eq:PVZ}
\notag \\
\end{eqnarray}
Above, 
$G_F$ 
denotes the Fermi coupling constant,  
$G_F = \sqrt{2} g^2  / 8 M_W^2$, 
and the subscripts
$V$
and
$A$
represent the vector and axial-vector quark bilinears, 
$\ol \psi \gamma_\mu \psi$ 
and
$\ol \psi \gamma_\mu \gamma_5$,
respectively. 
Notice we use a subscript on 
$\cL^{(5)}$
 to denote that both the coefficients and operators are taken at the scale
$\mu = M_Z$.

The operators appearing in 
$\mathcal{L}^{(5)}_{\mu = M_Z}$
are not closed under QCD renormalization, 
and at scales 
$\mu < M_Z$
other operators will be radiatively generated. 
A complete set of such operators has been enumerated in Sec.~\ref{s:OB}.
One should note that the non-penguin operators in 
Eq.~\eqref{eq:PVZ}  
come in exactly the combination
$A \otimes V + V \otimes A$.
In the chiral basis, 
this combination is proportional to the operator structure
$V_L \otimes V_L - V_R \otimes V_R$,
where 
$V_{L,R}$ 
refers to the fermion bilinear 
$( \ol \psi \gamma_\mu \psi )_{L,R}$.  
The orthogonal combination of operators has the structure
$V_L \otimes V_R - V_R \otimes V_L$. 
Non-penguin operators of the latter type are not generated from QCD radiative corrections at two-loop order, 
which is borne in by the block-diagonal structure of 
$\Gamma_{\D Q}$ 
in 
Eq.~\eqref{eq:GDQ}
below. 
This feature will persist to all orders in regularization schemes that preserve chirality~\cite{Ciuchini:1997bw}. 
The operator basis we employ thus contains the minimal set of operators necessary to describe 
isovector parity violation. 
Compared to the basis chosen by~\cite{Dai:1991bx}, 
we have two fewer operators from 
$Z^0$ 
exchange in each effective theory. 
Taking into account the Fierz constraint, 
there are only five operators at hadronic scales.

While using an operator basis with unnecessary operators has no physical consequences,
the computation of additional matrix elements using lattice QCD can consequently be unnecessarily costly. 
There is no computational advantage to using  
$A \otimes V+ V \otimes A$
operators instead of each one individually, 
however,
as we know of no algebraic way to combine the two sets of required quark contractions. 
There is an advantage to using 
$V \otimes A$ 
and 
$A \otimes V$
operators over 
$V_L \otimes V_L - V_R \otimes V_R$
and 
$V_L \otimes V_R - V_R \otimes V_L$
operators;
because, 
in the latter case, 
parity violation arises from the difference of parity-violating hadronic matrix elements, 
which is presumably statistically noisier than in the former case. 
Our minimal operator basis is thus economical only for perturbative calculations.
We will present our final results for Wilson coefficients in the operator basis of~\cite{Kaplan:1992vj}, 
which is practical for lattice computations.

\subsubsection{One-Loop Corrections}

As we work at next-to-leading order, 
we must additionally consider the one-loop QCD corrections arising from integrating out the 
top quark and 
$Z^0$ 
boson. 
At this order, 
the Wilson coefficients are determined by comparing one-loop graphs in the full and effective theories.

In the full theory, 
the next-to-leading order isovector parity-violating Lagrangian can be written as
\begin{equation} \label{eq:Loop}
\cL_{\mu = M_Z}^{(5)}
= 
\frac{G_F}{\sqrt{2}}
\sum_{i=1}^6
A_i(M_Z) \,
\mathcal{O}^{(5)}_i (M_Z)
,\end{equation}
where the amplitudes
$A_i(M_Z)$
are computed up to one-loop order, 
and are thus of the form
\begin{equation}
A_i 
(M_Z)
= 
(c_0)_i
+ 
\frac{\alpha_s(M_Z)}{4\pi}
(c_1)_i
.\end{equation}
The 
$(c_0)_i$
are determined at tree level, 
and can be read off from Eq.~\eqref{eq:PVZ}, 
namely
\begin{eqnarray}
(c_0)_1 
&=&
(c_0)_3
=
-
\frac{1}{3} \sin^2 \theta_W 
= 
-
0.0771
,\notag \\
(c_0)_5 
&=&
- 1 +2  \sin^2 \theta_W
=
-
0.5376
\label{eq:Tree}
,\end{eqnarray}
with all others zero.
The 
$(c_1)_i$
arise from the one-loop corrections, 
as we detail below.

At the scale
$\mu = M_Z$, 
the effective theory must reproduce 
Eq.~\eqref{eq:Loop}
at one-loop order. 
This is achieved by specifying values for the Wilson coefficients at this scale. 
At tree level, 
the Wilson coefficients are obviously given by 
$C_i^{(5)} (M_Z)= (c_0)_i$. 
At one-loop order, 
however, 
these coefficients are renormalized. 
After regularization and subtraction, 
the finite renormalization in the effective theory will have the form
\footnote{
Technically to derive Eq.~\eqref{eq:Rdef}, 
the matrix 
$r$ 
is computed by talking certain matrix elements of the operators, 
$\langle \, \cO_j^{(5)} (\mu) \, \rangle$. 
One must then isolate terms that arise from the operator renormalization from those that arise from the particular external states chosen to compute the matrix elements. 
}
\begin{equation}
\cL
= 
\frac{G_F}{\sqrt{2}}
\sum_{i,j=1}^6
(c_0)_i
\left(
\delta_{ij} 
+ 
\frac{\a_s(M_Z)}{4 \pi}
r_{ij}
\right)
\mathcal{O}^{(5)}_j (M_Z)
\label{eq:Rdef}
.\end{equation}
To reproduce the full theory, 
we must have
\begin{equation}
C_i^{(5)}(M_Z)
=
(c_0)_i
+ 
\frac{\alpha_s(M_Z)}{4 \pi}
\left[
(c_1)_i
- 
\sum_j
(c_0)_j \,
r_{ji} 
\right]
\label{eq:initial}
.\end{equation}

%
%
\begin{figure}
\epsfig{file=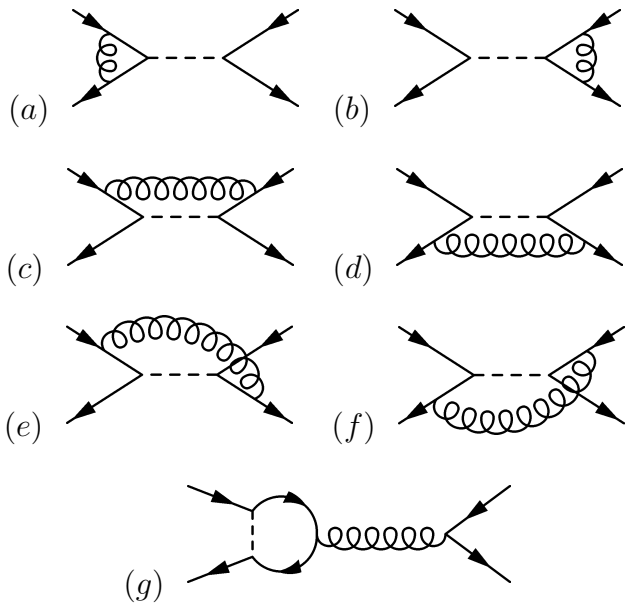,width=0.46\textwidth}
\caption{
One-loop diagrams in the QCD renormalization of $Z^0$ exchange.
The straight lines are quarks, 
while the curly lines are gluons, 
and dashed lines represent  $Z^0$ bosons. 
The wave-function renormalization contributes but is not depicted. 
}
\label{f:ZQCD}
\end{figure}
%
%

\begin{figure}
\epsfig{file=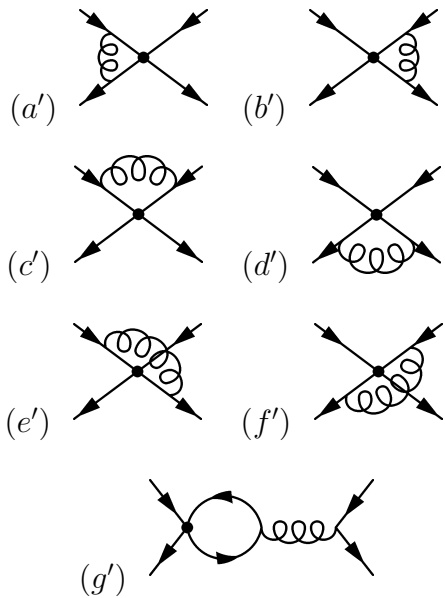,width=0.325\textwidth}
\caption{
One-loop diagrams in the QCD renormalization of parity violating four-quark operators.
Diagrams 
$(a')$--$(f')$ 
show the renormalization of current-current interactions, 
while diagram 
$(g')$ 
is a QCD penguin. 
Filled circles denote the four-quark operator,
while all other diagram elements are as in Fig.~\ref{f:ZQCD}. 
The wave-function renormalization contributes but is not depicted. 
}
\label{f:HPVclasses}
\end{figure}

To compute the Wilson coefficients at next-to-leading order, 
we must then determine the finite terms 
$(c_1)_i$ 
in the full theory, 
and the matrix 
$r$ 
in the effective theory. 
The former arises from the one-loop QCD corrections to $Z^0$ exchange, 
shown in Fig.~\ref{f:ZQCD},
whereas the latter arises from the one-loop corrections to the operators in 
Eq.~\eqref{eq:PVZ}, 
shown in Fig.~\ref{f:HPVclasses}.
The finite contributions to these diagrams are renormalization scheme dependent.

%
%
%
%

To perform these computations, 
we use dimensional regularization with 
$\ol {\text{MS}}$ 
subtraction in the 't Hooft--Veltman scheme for 
$\gamma_5$. 
Accordingly there are a total of 
$d = 4 - 2 \varepsilon$ 
matrices
$\gamma_\mu$, 
that satisfy 
$\{ \gamma_\mu, \gamma_\nu \} = 2 \, g_{\mu \nu}$, 
where 
$g_{\mu \nu}$
is the $d$-dimensional metric tensor.
The definition of the matrix 
$\gamma_5$
that appears in four-dimensions is retained in 
$d$-dimensions, 
so that 
$\gamma_5 =  - i \gamma_0 \gamma_1 \gamma_2 \gamma_3$.
Thus if we write the Dirac matrices in terms of 
four- and $(- 2 \varepsilon)$-components, 
$\gamma_\mu = \tilde{\gamma}_\mu + \hat{\gamma}_\mu$, 
then 
$\gamma_5$
anti-commutes in four-dimensions,
$\{ \tilde{\gamma}_\mu, \gamma_5 \} = 0$, 
and commutes in the remaining dimensions, 
$[ \hat{\gamma}_\mu, \gamma_5 ]= 0$. 
In extending the definition of chiral bilinears of the form
$(\ol \psi \gamma_\mu \psi )_{L,R}$
to 
$d$-dimensions, 
we must use the symmetric projection 
\begin{equation}
(\ol \psi  \gamma_\mu \psi)_{L,R}
\equiv
\ol \psi \cP_{R,L} \gamma_\mu \cP_{L,R} \psi
=
\ol \psi \tilde{\gamma}_\mu \cP_{L,R} \psi
\label{eq:leftright}
.\end{equation}

Having spelled out almost all the details of our renormalization scheme, 
we now proceed to match the full theory onto the effective theory at one-loop order in 
$\a_s$. 
In performing this matching, 
it is trivial to integrate out the top quark, 
as there are no diagrams with an internal top quark appearing at this order. 
There is a top quark loop in the 
$Z^0$ 
self-energy, 
however, 
this contribution is automatically taken into account by using the physical value for the mass
$M_Z$.

Ultraviolet divergences present in the one-loop QCD corrections to the current vertices
[diagrams $(a)$ and $(b)$ of Fig.~\ref{f:ZQCD}]
are exactly cancelled by the wave-function renormalization. 
Ordinarily finite contributions are additionally cancelled; 
however, 
in the 
$\ol {\text{MS}}$ 
subtraction scheme with 
't Hooft--Veltman regularization, 
the axial-current vertex receives a finite renormalization%
~\cite{Buras:1989xd}. 
Specifically the renormalized axial-vector current has the form
\begin{equation} \label{eq:shit}
A_\mu 
= 
\left( 1 + \frac{\alpha_s}{\pi} C_F \right)
\left(
\ol \psi_1
\tilde{\gamma}_\mu \gamma_5
\psi_2
\right)
,\end{equation}
with 
$C_F = \frac{1}{2} ( N_c^2 - 1 ) / N_c$, 
and for 
\emph{any} 
two quarks, 
$\psi_1$
and
$\psi_2$.
Such renormalization implies a non-vanishing two-loop anomalous dimension for the axial-vector current in this scheme. 
Because the operators in 
Eq.~\eqref{eq:PVZ} 
are each products of color-singlet bilinears, 
a similar fate befalls the analogous vertex corrections in the effective theory
[shown as diagrams $(a')$ and $(b')$ in Fig.~\ref{f:HPVclasses}].
The ultraviolet divergences in the effective theory are cancelled by wave-function renormalization, 
but finite contributions are left. 
The difference of these finite contributions between the full and effective theories vanishes.
In this scheme,
there is no 
$\cO(\a_s)$
contribution in 
Eq.~\eqref{eq:initial} 
from matching the one-loop current vertex diagrams. 
This scheme, 
however, 
has undesirable features, 
such as the mixing of left- and right-handed components due to Eq.~\eqref{eq:shit}, 
and the appearance of the axial-vector current anomalous dimension in the evolution of the Wilson coefficients. 
To preserve chirality, 
we augment the 't Hooft--Veltman scheme by imposing a multiplicative renormalization on the axial-vector current. 
The multiplicative factor is chosen to force the renormalized axial-vector current to have the form
\begin{equation}
A_\mu 
= 
\left(
\ol \psi_1
\tilde{\gamma}_\mu \gamma_5
\psi_2
\right)
+ 
\cO(\a_s^2)
.\end{equation}  
This augmented scheme is commonly employed to preserve chirality, 
see, e.g.,~\cite{Buras:1991jm,Ciuchini:1993vr}.
To impose the augmented scheme, 
we recompute the current-vertex diagrams in the full theory with the additional multiplicative renormalization. 
The next-to-leading order contributions from diagrams
$(a)$ 
and 
$(b)$
in 
Fig.~\ref{f:ZQCD} accordingly vanish. 
Diagrams 
$(a')$
and 
$(b')$
in the effective theory now give non-vanishing contributions to the Wilson coefficients,
upon taking the difference in 
Eq.~\eqref{eq:initial}.

In matching the penguin contributions in the full and effective theories, 
diagrams 
$(g)$ 
and 
$(g^\prime)$ 
of Figs.~\ref{f:ZQCD} and \ref{f:HPVclasses}, 
respectively, 
we note that only the operator
$\cO^{(5)}_{1}$
has a non-vanishing penguin contraction in the effective theory. 
The vanishing of penguin contractions for the other operators can be easily demonstrated. 
For the operator
$\cO^{(5)}_{5}$, 
the color-singlet 
$\ol b b$
cannot mix with the gluon;
and, 
for the operator 
$\cO^{(5)}_{3}$,
the left- and right-handed field cannot be contracted at one-loop order
[an all orders argument is given after Eq.~\eqref{eq:NoPeng} below].  
Carefully evaluating the penguin diagrams in the full and effective theory, 
we find exactly the same contractions are involved as the penguins considered in%
~\cite{Buras:1991jm}.
In 't Hooft--Veltman regularization, 
moreover,
the two diagrams are described by the same function, 
and their difference exactly cancels when the scale 
$\mu = M_Z$
is specified in the effective theory. 
The remaining next-to-leading order contributions to 
Eq.~\eqref{eq:initial} 
arise from comparing diagrams 
$(c)$--$(f)$
in the full theory, 
to 
$(c^\prime)$--$(f^\prime)$
in the effective theory. 
The former diagrams are finite, 
while the latter require regularization and subtraction. 
Identifying the renormalization scale of the latter diagrams with 
$M_Z$, 
moreover,
ensures that there are no logarithms in the matching conditions, 
Eq.~\eqref{eq:initial}.

Complete one-loop results for diagrams 
$(a)$--$(f)$ 
in the full theory are described by
$(c_1)_i$, 
whose non-zero entries are given by 
\begin{eqnarray}
(c_1)_1
&=&
- \frac{1}{3} 
(c_1)_2 
=
+ \frac{1}{2} (c_0)_1
,\notag \\
(c_1)_3
&=&
- \frac{1}{3} 
(c_1)_4 
=
-
\frac{1}{2} (c_0)_3
,\notag \\
(c_1)_5
&=&
- \frac{1}{3}
(c_1)_6
= 
+ \frac{1}{2}
 (c_0)_5
\label{eq:FULL}
.\end{eqnarray}
This result includes the multiplicative renormalization of the axial-vector current. 
In the effective theory, 
the finite contributions of diagrams
$(a')$--$(f')$ 
in Fig.~\ref{f:HPVclasses} are described by the matrix 
$r$
in Eq.~\eqref{eq:Rdef}.
The relevant rows of 
$r$ 
required for matching are
\begin{eqnarray}
r_{1j} 
&=&
\begin{pmatrix}
7 & -5 & 0 & 0 & 0 & \phantom{-}0 
\end{pmatrix}_j
,\notag \\
r_{3j}
&=&
\begin{pmatrix}
0 & \phantom{-}0 &  3 & 7 & 0 & \phantom{-}0 
\end{pmatrix}_j
,\notag \\
r_{5j}
&=&
\begin{pmatrix}
0 & \phantom{-}0 & 0 & 0 & 7 & -5
\end{pmatrix}_j
.\end{eqnarray}

Combining the results for 
$(c_1)_i$
and 
$r$, 
we can match the two theories at 
$\mu = M_z$, 
and deduce the Wilson coefficients, 
$C_i^{(5)}(M_Z)$.
We find
\begin{eqnarray}
C_1^{(5)} (M_Z) 
&=&
\left[ 1 - \frac{13}{2} \frac{\a_s(M_Z)}{4 \pi} \right]
(c_0)_1
,\notag\\
C_2^{(5)} (M_Z) 
&=&
\phantom{-}
\frac{7}{2}
\frac{\a_s(M_Z)}{4 \pi}
(c_0)_1 
,\notag\\
C_3^{(5)} (M_Z) 
&=&
\left[ 1 - \frac{7}{2} \frac{\a_s(M_Z)}{4 \pi} \right](c_0)_1
,\notag\\
C_4^{(5)} (M_Z) 
&=&
-
\frac{11}{2}
\frac{\a_s(M_Z)}{4 \pi} 
(c_0)_1
,\notag\\
C_5^{(5)} (M_Z) 
&=&
\left[ 1 - \frac{13}{2} \frac{\a_s(M_Z)}{4 \pi} \right]
(c_0)_5,
\notag\\
C_6^{(5)} (M_Z) 
&=&
\phantom{-}
\frac{7}{2}
\frac{\a_s(M_Z)}{4 \pi} 
(c_0)_5
\label{eq:ZMatch}
.\end{eqnarray}
The next-to-leading order corrections give rise to a
$\sim 5\%$
contribution to the input values of the Wilson coefficients.

\subsection{Isovector Parity-Violating Charged Current}
\label{s:PVCC}

The charged-current interaction also makes contributions to isovector hadronic parity violation. 
The non-leptonic charged current has the form
\begin{equation}
J_\mu^{W^-}
=
\frac{1}{\sqrt{2}}
\left( 
\ol U
\gamma_\mu V
D
\right)_L
,\end{equation}
with 
$J_\mu^{W^+} = (J_\mu^{W^-})^\dagger$.
The unitary matrix 
$V$ 
is the CKM matrix
(which should not be confused with the subscript 
$V$ 
denoting the vector spinor contraction). 
At scales below 
$M_W$, 
we can integrate out the 
top quark and 
$W$-bosons simultaneously. 
At tree level, 
the resulting five-flavor isovector parity-violating effective Lagrangian 
has the form
\begin{eqnarray}
\mathcal{L}_{\mu = M_W}^{\prime \; (5)}
&=& 
\frac{2 G_F}{\sqrt{2}}
\Bigg[
\left( 
\ol U
\gamma_\mu
V
D
\right)_L
\left( 
\ol D 
\gamma^\mu
V^\dagger
U
\right)_L
-
\{ L \leftrightarrow R \}
\Bigg]
.\notag\\
\label{eq:Wtree}
\end{eqnarray}
Here the top quark has been removed from the spinor
$U$, 
and the subscript on 
$\cL^{\prime \, (5)}$
is used to denote that both the coefficients and operators are taken at the scale 
$\mu = M_W$.

At this point, 
it is convenient to use a Fierz transformation. 
Such intrinsically four-dimensional identities are potentially problematic in 
$d$-dimensional renormalization schemes. 
In 't Hooft--Veltman dimensional regularization, 
the renormalization of operators is independent of whether they are Fierz transformed~\cite{Buras:1992tc}. 
After a Fierz transformation, 
the charged-current contribution to the isovector parity-violating Lagrangian has the form
\begin{eqnarray}
\mathcal{L}_{\mu = M_W}^{\prime \; (5)}
&=&
\frac{G_F}{\sqrt{2}}
\Bigg[
 |V_{us}|^2
( \ol u \gamma_\mu u - \ol d \gamma_\mu d \, ]_L
[ \,  \ol s \gamma^\mu s )_L
\notag \\
&&
-
|V_{cd}|^2
(\ol u \gamma_\mu u - \ol d \gamma_\mu d \, ]_L
[ \, \ol c \gamma^\mu c )_L
- 
\{ L \leftrightarrow R \}
\Bigg]
.\notag \\
\end{eqnarray}
In writing the effective Lagrangian above, 
we have dropped bottom quark contributions. 
This is because they enter proportional to 
$|V_{ub}|^2$, 
and the ratio
$|V_{ub}|^2 / |V_{us}|^2 \approx 2 \times 10^{-4}$ 
is negligible. 
Given the experimental values~\cite{Nakamura:2010zzi} for 
$|V_{us}|$ and $|V_{cd}|$, 
we can make a further approximation. 
With 
$|V_{us}| = 0.2252(9)$,
and 
$|V_{cd}| = 0.230(11)$, 
we shall take
$|V_{cd}| \approx |V_{us}|$. 
This approximation is quite reasonable because the charged-current 
component of isovector hadronic parity violation is only a 
$10 \%$ 
contribution compared to the neutral-current component. 
Dropping the bottom quark operators and treating 
$|V_{cd}| = |V_{us}|$
amount to neglecting sub-percent effects overall.

Under these numerical approximations, 
we have the charged-current contribution to 
$|\Delta I | = 1$
parity violation in the following form
\begin{eqnarray}
\mathcal{L}_{\mu = M_W}^{\prime \; (5)}
&=&
\frac{G_F |V_{us}|^2 }{\sqrt{2}}
\Bigg[
(\ol u \gamma_\mu u - \ol d \gamma_\mu d \, ]_L
[ \, \ol s \gamma^\mu s - \ol c \gamma^\mu c )_L
\notag \\
&& \phantom{space}
- \{ L \leftrightarrow R \}
\Bigg] \label{eq:WPV}
.\end{eqnarray}
While the operator in 
Eq.~\eqref{eq:WPV} 
does not mix with those in 
$\mathcal{L}_{\mu = M_Z}^{(5)}$
under QCD renormalization, 
an additional operator is required to form a closed set of charged-current contributions.
The closed set has been given above in Eq.~\eqref{eq:5B}.  
From comparing
Eqs.~\eqref{eq:WPV} and \eqref{eq:5B}, 
moreover, 
we find the tree-level values of the remaining Wilson coefficients in the five-flavor effective theory, 
\begin{eqnarray}
C_{7}^{(5)}(M_W) 
&=& 
0
\notag \\
C_{8}^{(5)}(M_W) 
&=&
|V_{us}|^2
=
- 0.0508
\label{eq:WMatch}
.\end{eqnarray}

Finally as the charged-current interaction leads to a subdominant contribution to 
$|\Delta I | = 1$ 
hadronic parity violation, 
we do not consider its one-loop QCD corrections.
Omitting such corrections amounts to neglecting sub-percent effects, 
which is consistent with the approximations used thus far. 
The tree-level values for Wilson coefficients will accordingly be used for charged-current contributions.

\section{Two-Loop Running} \label{s:RUN}%

\subsection{QCD Renormalization Group} \label{ss:RGE}  %

\subsubsection{Evolution of Wilson Coefficients}

The Wilson coefficients, 
$C_i^{(N_f)}(\mu)$ 
evolve with the scale 
$\mu$
according to the renormalization group equation, 
\begin{equation} \label{eq:CWrun}
\mu \frac{d}{d \mu} C_i^{(N_f)}
=
\sum_{j}
C_j^{(N_f)} \,
\Gamma_{ji}^{(N_f)} (\alpha_s)
,\end{equation}
with 
$\Gamma^{(N_f)}$
as the anomalous dimension matrix in the theory with 
$N_f$
flavors. 
The entry 
$\Gamma^{(N_f)}_{ii}$ 
describes the renormalization of the operator
$\mathcal{O}_i^{(N_f)}$, 
while 
$\Gamma^{(N_f)}_{ji}$
describes its mixing into the operator 
$\mathcal{O}_j^{(N_f)}$. 
It is convenient to drop the superscript 
$(N_f)$
for notational ease. 
When we match between different effective theories, 
the superscript will reappear.

At next-to-leading order, 
the anomalous dimension matrix can be written as
\begin{equation}
\Gamma (\alpha_s)
= 
\frac{\alpha_s}{4 \pi}
\gamma_0
+
\left( \frac{\alpha_s}{4 \pi} \right)^2
\gamma_1
,\end{equation}
with the dependence on 
$\alpha_s$ 
made explicit.
Here 
$\gamma_0$
is the leading-order anomalous dimension matrix, 
and 
$\gamma_1$
is the next-to-leading order matrix. 
The QCD running coupling satisfies the renormalization group equation
\begin{equation}
\mu^2 \frac{d}{d \mu^2} 
\alpha_s
=
- \beta_0 
\frac{\alpha_s^2}{4 \pi}
- \beta_1 
\frac{\alpha_s^3}{(4\pi)^2}
,\end{equation}
with 
$\beta_0 = \frac{1}{3} (11 N_c - 2 N_f)$, 
and 
$\beta_1 = \frac{34}{3} N_c^2 - \frac{10}{3} N_c N_f -  (N_c - N_c^{-1}) N_f$, 
the leading and next-to-leading order coefficients of the 
$\beta$-function, 
respectively.  
The solution to the renormalization group equation for 
$\alpha_s (\mu)$ 
can be written as
\begin{equation} \label{eq:Alpha}
\alpha_s(\mu)
= 
\frac{4 \pi}{\beta_0 \log ( \mu^2 / \Lambda^2 )}
\left[
1 - \frac{\beta_1}{\beta^2_0}
\frac{\log \log ( \mu^2 / \L^2)}{\log ( \mu^2 / \L^2)}
\right]
,\end{equation}
which serves to define the QCD scale 
$\Lambda$.

The solution to the renormalization group equation for the Wilson coefficients, 
Eq.~\eqref{eq:CWrun}, 
can be written as
\begin{equation} \label{eq:Soln}
C_i (\mu^\prime)
=
\sum_j
U_{ij}(\mu',\mu) 
C_j (\mu)
.\end{equation}
Here 
$U(\mu^\prime, \mu)$
is the evolution matrix. 
Valid to next-to-leading order, 
the explicit form of the evolution matrix is specified by~\cite{Buras:1991jm}
\begin{equation}
U(\mu^\prime,\mu)
=
\left( 1 + \frac{\alpha_s(\mu^\prime)}{4 \pi} J \right)
U_0 (\mu^\prime, \mu)
\left( 1 - \frac{\alpha_s(\mu)}{4 \pi} J \right)
,\end{equation}
with 
$U_0$
as the leading-order evolution matrix
\begin{equation}
U_0(\mu^\prime, \mu)
=
\mathcal{V}
\left(
\frac{\alpha_s(\mu^\prime)}{\alpha_s(\mu)}
\right)^{-\frac{ (\gamma_{0})_\mathcal{D}}{2 \beta_0}}
\mathcal{V}^{-1}
,\end{equation}
defined in terms of the matrix 
$\mathcal{V}$
that diagonalizes 
$\gamma_0^T$, 
namely
$(\gamma_0)_\mathcal{D} = \mathcal{V}^{-1} \gamma^T_0 \mathcal{V}$. 
The matrix 
$J$
appears in the form
\begin{equation}
J
= 
\gamma_0^T
\frac{\beta_1}{2 \beta_0^2}
-
\mathcal{V}
K
\mathcal{V}^{-1}
,\end{equation}
where the matrix 
$K$
is defined by
\begin{equation}
K_{ij}
=
\frac{ [\mathcal{V}^{-1} \gamma^T_1 \mathcal{V}]_{ij}}
{2 \beta_0 + [ ( \gamma_0)_\mathcal{D}]_i - [ ( \gamma_0)_\mathcal{D}]_j}
\label{eq:K}
.\end{equation}

\subsubsection{Matching at Heavy Quark Thresholds}

The evolution of Wilson coefficients expressed in Eq.~\eqref{eq:Soln} is applicable as we run down 
from,
say, 
$\mu = M_Z$ 
to 
$\mu = m_b$. 
In order to evolve to lower scales, 
we must confront the matching from a five-flavor to a four-flavor effective theory,
or, 
more generally, 
from an 
$N_f$-flavor 
to an 
$(N_f -1)$-flavor 
effective theory.
Matching of theories at the heavy quark threshold has two parts:
continuity of the coupling constant, 
and 
determination of the Wilson coefficients, 
$C_i^{(N_f -1)}(m_Q)$.

As the QCD 
$\b$-function is 
$N_f$ dependent, 
the coupling, 
$\a_s(\mu)$, 
will be discontinuous as we pass through a quark threshold, 
$\mu = m_Q$. 
This unphysical discontinuity is avoided by adjusting the QCD scale parameter 
$\Lambda$ 
in Eq.~\eqref{eq:Alpha}. 
For a theory with 
$N_f$ 
flavors, 
the QCD scale is then
$\Lambda^{(N_f)}$, 
with continuity in the coupling enforced at the quark threshold, 
$m_Q$, 
by fixing  
$\Lambda^{(N_f -1)}$ 
so that
\begin{equation}
\alpha_s^{(N_f)} (m_Q, \Lambda^{(N_f)}) 
= 
\alpha_s^{(N_f-1)} (m_Q, \Lambda^{(N_f -1)})
.\end{equation}

Having fixed the QCD coupling to be continuous, 
we must determine values for the Wilson coefficients, 
$C_i^{(N_f -1)}(m_Q)$, 
from the 
$C_i^{(N_f)}(m_Q)$.
In order to match the 
$N_f$- 
and 
$(N_f - 1)$-flavored 
theories at the threshold
$\mu = m_Q$, 
we must first regroup the set of operators
$\{ \cO_i^{(N_f)} \}$
to match those in the 
$(N_f -1)$-flavor theory when the heavy quark is integrated out. 
It is best to spell this out explicitly for matching at the two scales
$\mu = m_b$, 
and 
$\mu = m_c$.

To match at the bottom quark threshold, 
we note that for 
$N_f = 5$, 
there are eight operators in the effective theory, 
and only six operators in the four-flavor theory.  
The third set of operators, 
$\cO_{7,8}^{(5)} (\mu)$,
in Eq.~\eqref{eq:5B}
is already present in the four-flavor theory, 
namely as
$\cO_{5,6}^{(4)} (\mu)$
in Eq.~\eqref{eq:PVC2}.
As we integrate out the bottom quark, 
the values of the 
$C^{(5)}_{7,8} (m_b)$
coefficients will smoothly go over as contributions to the
$C^{(4)}_{5,6} (m_b)$. 
The remaining six operators require matching conditions to avoid discontinuities when the bottom quark is integrated out.

We now spell this matching out explicitly. 
At the bottom quark threshold, 
$\mu = m_b$, 
we integrate out the bottom quark at one-loop order in the five-flavor effective theory. 
This produces four-flavor operators:
$\cO^{(5)}_{1\text{--}4} (m_b) \to \cO^{(4)}_{1\text{--}4} (m_b)$, 
$\cO^{(5)}_{5,6}(m_b) \to \cO^{(4)}_{5,6}(m_b)$, 
and
$\cO^{(5)}_{7,8} (m_b) = \cO^{(4)}_{5,6} (m_b)$.  
After regularization and subtraction, 
the one-loop amplitudes are described by
$\cO(\a_s)$
finite terms encoded in the matrix
$r^{(5)}$. 
In a theory with 
$N_f$
flavors, 
the matrix  
$r^{(N_f)}$
takes into account operator mixing at next-to-leading order, 
similar to 
$r$ 
in Eq.~\eqref{eq:Rdef}.  
Here,
the matrix
$r^{(N_f)}$ 
arises from the finite contributions to the one-loop renormalization of 
\emph{all} 
four-quark operators.

Because the set of operators 
$\cO^{(5)}_{7,8} (\mu)$
in 
Eq.~\eqref{eq:5B} 
is closed under renormalization, 
the full matrix
$r^{(5)}$
must have a block diagonal structure. 
The first block is an 
$6 \times 6$ 
dimensional matrix, 
while the second block is a 
$2 \times 2$ 
dimensional matrix. 
The latter matrix is identical to 
$r^{(4)}_{ij}$, 
for 
$i,j = 5$ 
or 
$6$,
because the corresponding operators are identical, 
and there is no effect at one-loop from integrating out the bottom quark. 
In the five-flavor theory, 
the effective action at 
$\mu = m_b$ 
is thus proportional to
\begin{eqnarray}
\sum_{i,j=1}^6
&& 
C^{(5)}_i (m_b) 
\left[
\delta_{ij} 
+ 
\frac{\a_s(m_b)}{4 \pi}
r^{(5)}_{ij}
\right] 
\mathcal{O}_j^{(4)} (m_b)
\notag  \\
&&+ 
\sum_{i,j=1}^2
C^{(5)}_{i+6} (m_b) 
\left[
\delta_{ij} 
+ 
\frac{\a_s(m_b)}{4 \pi}
r^{(4)}_{i+4,j+4}
\right] 
\mathcal{O}_{j+4}^{(4)} (m_b)
,\notag 
\\
\label{eq:MESS}
\end{eqnarray}
while in the four-flavor theory, 
the next-to-leading order effective action at 
$\mu = m_b$
is proportional to
\begin{eqnarray}
\sum_{i,j=1}^6 C^{(4)}_i (m_b) 
\left[
\delta_{ij}
+ 
\frac{\a_s(m_b)}{4 \pi}
r^{(4)}_{ij}
\right]
\mathcal{O}_j^{(4)} (m_b)
\label{eq:match}
,\end{eqnarray}
with the same constant of proportionality. 
In order that the theories match, 
we must have
\begin{align}
C_i^{(4)} (m_b) 
=
\sum_{j=1}^6
C_j^{(5)} (m_b)
\left[
\delta_{ji} + \frac{\a_s(m_b) }{4 \pi}
\Delta r^{(5)}_{ji} 
\right]
\notag \\
+
(\delta_{i5} + \d_{i6} ) \,
C_{i+2}^{(5)}(m_b)
\label{eq:Bmatch}
,\end{align}
where, 
in general, 
we have defined
\begin{equation} 
\Delta r^{(N_f)} \equiv  r^{(N_f)} - r^{(N_f - 1)}
\label{eq:Dr}
.\end{equation}
Notice that to perform the matching at threshold, 
it is only the difference of the flavor-dependent mixing matrices, 
Eq.~\eqref{eq:Dr}, 
that is required. 
This difference incorporates the finite one-loop contributions from integrating out the heavy quark.

Matching at the charm quark threshold is simpler. 
At the scale 
$\mu = m_c$, 
we integrate out the charm quark, 
and the operators 
$\cO^{(4)}_i (m_c)$
become 
$\cO^{(3)}_i (m_c)$,
for each 
$i= 1, \ldots, 6$. 
This fact enables us to determine the matching condition, 
\begin{equation}
C_i^{(3)} (m_c)
=
\sum_{j,k}
C_j^{(4)} (m_c)
\left[
\delta_{jk}
+
\frac{\a_s (m_c)}{4 \pi}
\D r^{(4)}_{jk}
\right]
\label{eq:Cmatch}
,\end{equation}
with the matrix 
$\D r^{(4)}$
as the difference of finite contributions in the two effective theories,
given in  
Eq.~\eqref{eq:Dr}.

\subsubsection{Solution: Wilson Coefficients at Hadronic Scales}

To summarize this section, 
we reiterate how to determine the Wilson coefficients at hadronic scales, 
$\Lambda_{\text{QCD}} < \mu < m_c$.
The input values are 
$C^{(5)}_{1\text{--}6}(M_Z)$, 
and
$C^{(5)}_{7,8}(M_W)$
determined above in Sec.~\ref{s:C}. 
These values are evolved down to a common scale 
$\mu = m_b$, 
using the next-to-leading order evolution in Eq.~\eqref{eq:Soln}.
Fortunately these two sets of input values can be run down independently because they do not mix under renormalization. 
From the eight values 
$C^{(5)}_i(m_b)$, 
we match at the bottom quark threshold, 
via Eq.~\eqref{eq:Bmatch}, 
to determine the six Wilson coefficients in the four-flavor effective theory, 
$C^{(4)}_i (m_b)$. 
These we run down to the charm quark mass, 
\begin{equation}
C^{(4)}_i (m_c)
= 
\sum_{j}
U_{ij}^{(4)} (m_c, m_b)
C_{j}^{(4)} (m_b)
.\end{equation}
Matching at 
$\mu = m_c$
is accomplished using 
Eq.~\eqref{eq:Cmatch}
to determine
$C_i^{(3)}(m_c)$. 
The desired Wilson coefficients, 
$C_i^{(3)}(\mu)$, 
for 
$\Lambda_{\text{QCD}} < \mu < m_c$, 
are then simply
\begin{equation}
C_i^{(3)}(\mu)
= 
\sum_j
U_{ij}^{(3)} (\mu, m_c) 
C_j^{(3)}(m_c)
.\end{equation}
Finally to perform the evolution from 
the weak scale down to hadronic scales, 
we need to know the anomalous dimension matrices
$\Gamma^{(N_f)}$
to perform the running, 
and the difference of mixing matrices
$\D r^{(N_f)}$ 
to perform the matching.
These quantities are discussed next.

\subsection{Anomalous Dimensions} \label{eq:ADM}                                       %

The operators, 
$\mathcal{O}_i^{(N_f)}$,
are renormalized by gluon radiation. 
There are two generic classes of Feynman diagrams according 
to which the gluon radiation can be classified. 
These diagram classes are current-current interactions, 
and QCD penguins. 
One-loop diagrams are depicted in Fig.~\ref{f:HPVclasses}.
At two-loop order, 
one attaches an additional gluon in all possible places, 
and the diagrams can again be classified in these two categories.

Current-current renormalization does not alter the flavor structure of the operator in question.
QCD penguins, 
however,  
can lead to the generation of so-called penguin operators that contain
the flavor-singlet combination proportional to 
$\sum_q (\ol q q)_V$.
At next-to-leading order, 
QCD radiation can also generate operators that contain
$\sum_q (\ol q q)_A$. 
There are two species of penguins, 
moreover, 
depending on how the quarks in the operator are contracted. 
Type-one penguins 
are formed from a closed quark loop, 
i.e.~when spin contracted quarks, 
such as 
$\psi_3$ 
and 
$\ol \psi_3$
in  
$\mathcal{O} = (\ol \psi_1 \gamma_\mu \psi_2)_L (\ol \psi_3 \gamma^\mu \psi_3)_L$,  
are Wick contracted along with gluon interactions. 
Type-two penguins, 
on the other hand, 
are formed when quarks with uncoupled spin are Wick contracted along with gluon interactions, 
such as 
$\psi_3$ and $\ol \psi_3$ in 
$\mathcal{O}^\prime = (\ol \psi_1 \gamma_\mu \psi_3)_L (\ol \psi_3 \gamma^\mu \psi_2)_L$. 
It will be helpful to keep the two classes of diagrams and two types of penguins in mind 
when discussing the renormalization. 
A thorough discussion of renormalization of four-quark operators at next-to-leading order is given in~\cite{Buchalla:1995vs}.

\subsubsection{$N_f = 5$}%

Let us begin with the five-flavor effective theory valid at scales above the bottom quark mass. 
The operators 
$\mathcal{O}_i^{(5)}$
are given in Eqs.~\eqref{eq:5}, \eqref{eq:5A}, and \eqref{eq:5B}. 
In this theory, 
the anomalous dimension matrix, 
$\Gamma^{(5)}$,
is best considered in blocks corresponding to these three sets of operators. 
As current-current renormalization can alter the color structure of operators, 
but not the flavor structure, 
each block is closed under renormalization from current-current type gluon radiation. 
Due to their flavor structure, 
operators in the last block are not renormalized by QCD penguin diagrams. 
Thus these operators do not mix with the other two blocks. 
Operators in the second block can mix with the first block through penguin diagrams of type one. 
Finally operators in the first block are renormalized by both
type one and two penguin diagrams, 
however, 
as these are the penguin operators themselves, 
their flavor structure is not altered outside their own block. 
Based on these general considerations, 
the anomalous dimension matrix must have the block form
\begin{equation}
\Gamma^{(5)}
=
\begin{pmatrix}
\cC+ 5 P + 2 P^\prime & 0 & 0 \\
P_{2 \times 4} & \cC_{2 \times 2} & 0 \\
0 & 0 & \cC_{2 \times 2}  
\end{pmatrix}
\label{eq:ADM5}
,\end{equation}
where the 
$4 \times 4$
blocks 
$\cC$, 
$P$, 
and 
$P^\prime$
are the anomalous dimension matrices from 
current-current renormalization and penguins 
of type one and two, respectively. 
The matrix 
$\cC_{2 \times 2}$, 
which encodes the anomalous dimensions of current-current radiation of 
$V_L \otimes V_L - V_R \otimes V_R$ 
operators, 
is the upper-left 
$2 \times 2$ 
submatrix of 
$\cC$. 
The matrix 
$P_{2 \times 4}$
consists of the top 
$2 \times 4$ 
submatrix of
$P$
and encodes the type-one pengiun radiation from 
$V_L \otimes V_L - V_R \otimes V_R$
operators. 
Each of these matrices has an expansion in 
$\alpha_s$, 
of which we consider only the first two terms, 
\begin{eqnarray}
\cC &=& \frac{\alpha_s}{4 \pi} \cC_0 + \left( \frac{\a_s}{4 \pi} \right)^2 \cC_1,
\notag \\
P &=& \frac{\alpha_s}{4 \pi} P_0 + \left( \frac{\a_s}{4 \pi} \right)^2 P_1,
\notag \\
P^\prime &=& \frac{\alpha_s}{4 \pi} P^\prime_0 + \left( \frac{\a_s}{4 \pi} \right)^2 P^\prime_1,
\end{eqnarray} 
so that 
$\gamma_0^{(N_f)}$ 
consists solely of the matrices 
$\cC_0$, $P_0$, and $P^\prime_0$
without any powers of $\alpha_s$, 
and 
$\gamma_1^{(N_f)}$
similarly consists only of 
$\cC_1$, $P_1$, and $P^\prime_1$. 
Furthermore as these matrices appear in the other effective theories, 
we postpone giving the matrix elements until we have first discussed the general form in all three theories.  
The numerical factors in front of penguins in Eq.~\eqref{eq:ADM5}
obviously arise from the number of flavor contractions that can produce such contributions. 
In writing Eq.~\eqref{eq:ADM5}, 
we have made considerable use of the flavor-blindness of gluon interactions.

\subsubsection{$N_f = 4$}%

In the four-flavor effective theory, 
the operators are enumerated in Eqs.~\eqref{eq:PVC1} and \eqref{eq:PVC2}. 
Treating these sets in blocks,
each block undergoes current-current renormalization, 
but only the first block of operators can mix with QCD penguin operators. 
The anomalous dimension matrix in the 
four-flavor theory thus has the block-diagonal form
\begin{equation}
\Gamma^{(4)}
=
\begin{pmatrix}
\cC + 4 P + 2 P^\prime & 0 \\
0 & \cC_{2 \times 2}  \\
\end{pmatrix}
\label{eq:ADM4}
,\end{equation}
where the matrices
$\cC$, 
$P$, 
and 
$P^\prime$
are the same ones that appear in Eq.~\eqref{eq:ADM5}.

\subsubsection{$N_f = 3$}%

Finally at hadronic scales, 
we have the three-flavor effective theory. 
The relevant operators have been given in Eqs.~\eqref{eq:3PV} and \eqref{eq:SPV}. 
It is similarly efficacious to organize the anomalous dimension 
matrix in blocks. 
Each block undergoes current-current renormalization as well as 
renormalization from penguin diagrams. 
As above, 
only the first block receives contributions from penguins of type two.
In terms of the matrices 
$\cC$, 
$P$, 
and
$P^\prime$
employed above, 
the form of the anomalous dimension matrix must be
\begin{equation}
\Gamma^{(3)}
=
\begin{pmatrix}
\cC + 3 P + 2 P^\prime & 0 \\
P_{2 \times 4} & \cC_{2 \times 2} \\
\end{pmatrix}
\label{eq:G3}
.\end{equation}

\subsection{Finite Mixing Matrices}

At next-to-leading order, 
one also requires the matrices 
$r^{(N_f)}$
that characterize finite contributions to operator mixing at one-loop order. 
In each of the three effective theories, 
$N_f = 5$, 
$4$,
and
$3$, 
the general structure of the matrix 
$r^{(N_f)}$ 
is obviously identical to that of the
$\Gamma^{(N_f)}$
deduced above.
For example, 
in the four-flavor effective theory, 
we have
\begin{equation}
r^{(4)}
= 
\begin{pmatrix}
r_\cC + 4 r_P + 2 r_{P^\prime} & 0 \\
0 & r_{\cC_{2 \times 2}} \\
\end{pmatrix}
,\end{equation} 
where the subscripts denote the type of gluon radiation that leads to these finite contributions. 
As these 
$r^{(N_f)}$
matrices are required to perform the matching of Wilson coefficients at heavy quark thresholds,
moreover, 
we only need their differences, 
$\D r^{(N_f)}$
defined in Eq.~\eqref{eq:Dr}.  
At one-loop order, 
the only dependence on the number of flavors is generated by the type-one penguin contraction,
see 
Eqs.~\eqref{eq:ADM5}, 
\eqref{eq:ADM4}, 
and 
\eqref{eq:G3}.  
Physically these are the only relevant diagrams because they contain a heavy quark loop.

In matching the reduced basis
($i = 1, \ldots, 6$) 
of five-flavor operators at the scale 
$\mu = m_b$ 
to the four-flavor operators via Eq.~\eqref{eq:Bmatch}, 
we have the 
$6 \times 6$
dimensional matrix
\begin{equation}
\D r^{(5)}
= 
\begin{pmatrix}
r_P & 0 \\
r_{P_{2 \times 4}} & 0 
\end{pmatrix}
.\end{equation}
The 
$4 \times 4$ 
block matrix 
$r_P$
contains the finite contributions from the one-loop mixing due to the type-one penguin contraction of the bottom quark, 
and 
$r_{P_{2 \times4}}$
is the upper 
$2 \times 4$
submatrix of 
$r_P$. 
In matching the four-flavor theory to the three-flavor theory at the charm quark threshold via Eq.~\eqref{eq:Cmatch}, 
we require the matrix
\begin{equation}
\D r^{(4)} 
= 
\begin{pmatrix}
\phantom{-} r_P & 0 \\
- r_{P_{2 \times 4}} & 0 
\end{pmatrix}
,\end{equation}
which contains same type of contributions; 
this time due to the charm quark loop. 
Thus to perform the matching at heavy quark thresholds, 
it is sufficient to know the matrix
$r_P$.

\subsection{Matrix Elements} \label{eq:MEL}                                                    %

We have seen that to specify the anomalous dimension matrices for each of the effective
field theories, 
we must determine the three matrices 
$\cC$, 
$P$ 
and 
$P^\prime$. 
We do this to next-to-leading order. 
Furthermore, 
to match the effective theories at heavy quark thresholds, 
we additionally need the matrix 
$r_P$
that arises from finite contributions to the heavy quark loops.

\subsubsection{Current-Current Anomalous Dimensions}

The matrix 
$\cC$
arises from renormalization of the current-current interaction. 
We use a mass-independent renormalization scheme, 
so that the quarks are massless in each effective theory.
Additionally the QCD radiation is flavor blind.
These two features allow us to determine
$\cC$ 
from the renormalization of operators with four distinguishable quarks, 
$\psi_1, \ldots, \psi_4$. 
In order to correspond to our 
$4 \times 4$ 
blocks, 
these operators are written as
\begin{eqnarray}
\D Q_1 
&=& 
(\ol \psi_1 \gamma_\mu  \psi_2)_{L} (\ol \psi_3 \gamma^\mu  \psi_4)_{L} 
-
\{ L \leftrightarrow R \}
,\notag \\
\D Q_2
&=&
(\ol \psi_1 \gamma_\mu  \psi_2 \, ]_{L} [ \, \ol \psi_3 \gamma^\mu  \psi_4)_{L} 
-
\{ L \leftrightarrow R \}
,\notag \\
\D Q_3
&=&
(\ol \psi_1 \gamma_\mu  \psi_2)_{L} (\ol \psi_3 \gamma^\mu  \psi_4)_{R} 
-
\{ L \leftrightarrow R \}
,\notag \\
\D Q_4 
&=&
(\ol \psi_1 \gamma_\mu  \psi_2 \, ]_{L} [ \, \ol \psi_3 \gamma^\mu  \psi_4)_{R} 
-
\{ L \leftrightarrow R \}
.\label{eq:DQ}
\end{eqnarray}
Denoting the anomalous dimension matrix of these operators as
$\Gamma_{\D Q}$, 
we have simply
$\cC = \Gamma_{\D Q}$.

Next we observe that parity transformed operators share the same anomalous dimension. 
This fact owes to the parity invariance of QCD, as well as that of the regularization scheme. 
Finally in the augmented 't Hooft--Veltman scheme at next-to-leading order, 
the 
$V_L \otimes V_L$ 
and 
$V_L \otimes V_R$ 
operators do not mix under renormalization.
Their anomalous dimension matrices, 
$\Gamma_{Q_{\text{VLL}}}$ 
and 
$\Gamma_{Q_{\text{VLR}}}$, 
can be identified from the results determined in~\cite{Buras:1992tc}.
In that work,  
care was taken to handle evanescent operators that appear at intermediate stages of the calculation. 
From these matrices, 
we form
\begin{equation}
\Gamma_{\Delta Q} 
= 
\begin{pmatrix}
\Gamma_{Q_{\text{VLL}}} & 0 \\
0 & \Gamma_{Q_{\text{VLR}}} \\
\end{pmatrix}
\label{eq:GDQ}
,\end{equation} 
and consequently find
\begin{equation}
\cC_{2 \times 2}  = \Gamma_{Q_{\text{VLL}}}
.\end{equation}

For completeness, 
the required anomalous dimension matrices for 
$V_L \otimes V_L$
operators are
\begin{equation}
(\gamma_{Q_{\text{VLL}}})_0
=
\begin{pmatrix}
-2 & \phantom{-}6 \\
\phantom{-}6 & -2
\end{pmatrix}
,\end{equation}
at leading order, 
and 
\begin{equation}
(\gamma_{Q_{\text{VLL}}})_1
=
\begin{pmatrix}
\frac{553}{6} - \frac{58}{9} N_f & \frac{95}{2} - 2 N_f \\
\\
\frac{95}{2} - 2 N_f & \frac{553}{6} - \frac{58}{9} N_f 
\end{pmatrix}
,\end{equation}
at next-to-leading order~\cite{Buras:1992tc}. 
As for the 
$V_L \otimes V_R$
operators, 
we have the anomalous dimension matrix
\begin{equation}
(\gamma_{Q_{\text{VLR}}})_0
=
\begin{pmatrix}
2 & -6 \\
0 & \phantom{-}16
\end{pmatrix}
,\end{equation}
at leading order, 
and the matrix
\begin{equation}
(\gamma_{Q_{\text{VLR}}})_1
=
\begin{pmatrix}
121 - \frac{62}{9} N_f & -39 - \frac{2}{3} N_f  
\\ 
\\
\frac{95}{2}  - \frac{4}{3} N_f & - \frac{85}{2} - \frac{44}{9} N_f
\end{pmatrix}
,\end{equation}
at next-to-leading order~\cite{Buras:1992tc}.

\subsubsection{Penguins of Type One}

Renormalization of QCD penguins of type one
leads to the anomalous dimension matrix 
$P$
employed above. 
This matrix can be determined by studying how four generic operators 
are renormalized by QCD penguins of the first type. 
Due to flavor blindness of the QCD interaction, 
these operators can be chosen as
\begin{equation} 
\D P_i 
= 
\Delta Q_i \Big|_{\psi_4 = \psi_3}
\label{eq:DP}
,\end{equation} 
with $\D Q_i$ shown in 
Eq.~\eqref{eq:DQ}.
Notice that by design, 
the 
$\D P_i$
operators only have penguin contractions of type one.

The renormalization of the chiral-basis operators, 
$\D P_i$,
from penguin diagrams can also be determined from the results of%
~\cite{Buras:1992tc}, 
owing to the parity invariance of QCD and of the regularization scheme.
To determine the matrix 
$P$, 
it is sufficient to study how the set of four operators
$\{ \Delta P_i \}$, 
mixes with the penguin operators, 
\begin{eqnarray}
\D \cP_1 
&=&
( \ol \psi_1 \gamma_\mu \psi_2)_L \sum_q^{N_f} (\ol q  \gamma^\mu q)_L
-
\{ L \leftrightarrow R \}
,\notag \\
\D \cP_2
&=&
( \ol \psi_1 \gamma_\mu \psi_2 ]_L \sum_q^{N_f} [ \, \ol q  \gamma^\mu q)_L
-
\{ L \leftrightarrow R \}
,\notag \\
\D \cP_3 
&=&
( \ol \psi_1 \gamma_\mu \psi_2)_L \sum_q^{N_f} (\ol q  \gamma^\mu q)_R
-
\{ L \leftrightarrow R \}
,\notag \\
\D \cP_4
&=&
( \ol \psi_1 \gamma_\mu \psi_2 ]_L \sum_q^{N_f} [ \, \ol q  \gamma^\mu q)_R
-
\{ L \leftrightarrow R \}
.\label{eq:P2Chi}
\end{eqnarray}
This mixing is described by a 
$4 \times 4$
matrix 
$\Gamma_{\D P}$, 
and our desired matrix is simply
$P = \Gamma_{\D P}$.

For completeness, 
the leading-order type-one penguin anomalous dimension matrix 
has the form
\begin{equation}
(\gamma_{\D P})_0
=
\begin{pmatrix}
\phantom{-}0 & 0 & \phantom{-}0 & 0 
\\ 
- \frac{2}{9} & \frac{2}{3} & - \frac{2}{9} & \frac{2}{3} 
\\ 
\phantom{-}0 & 0 & \phantom{-}0 & 0 
\\ 
- \frac{2}{9} & \frac{2}{3} & - \frac{2}{9} & \frac{2}{3} 
\end{pmatrix}
.\end{equation}
While at next-to-leading order, 
the required anomalous dimension matrix can be deduced from the results of~\cite{Buras:1992tc}.
We find
\begin{equation}
(\gamma_{\D P})_1
= 
\begin{pmatrix}
\phantom{-}
\frac{23}{3}
& 
1 
& 
- 
\frac{25}{3}
& 
1 
\\
\\
- \frac{418}{243}
& 
\frac{850}{81}
& 
-\frac{1210}{243}
& 
\frac{490}{81}
\\
\\ 
- \frac{73}{9} 
& 
\frac{1}{3} 
& 
\phantom{-}\frac{71}{9}
& 
\frac{1}{3} 
\\
\\
- \frac{1246}{243} 
& 
\frac{382}{81}
& 
- \frac{256}{243}
& 
\frac{832}{81}
\\
\end{pmatrix}
.\end{equation}
Considerable care was taken in~\cite{Buras:1992tc} to ensure that contributions from evanescent and off-shell operators were taken into account at intermediates stages of the calculation.

The final type-one penguin contributions we require are their finite contributions at one-loop. 
These contributions are contained in the matrix%
~\cite{Buras:1991jm}
\begin{equation}
r_{\D P}
= 
\begin{pmatrix}
0 & \phantom{-}0 & 0 & \phantom{-}0 \\
\frac{5}{27} & - \frac{5}{9} & \frac{5}{27} & - \frac{5}{9} \\
0 & \phantom{-}0 & 0 & \phantom{-}0 \\
\frac{5}{27} & - \frac{5}{9} & \frac{5}{27} & - \frac{5}{9} \\
\end{pmatrix}
.\end{equation}
As deduced above, 
this is the only matrix of finite contributions necessary to match effective theories at heavy quark thresholds.

\subsubsection{Penguins of Type Two}
\label{s:P2}

To deduce the form of the anomalous dimension matrix that encodes the mixing of operators due to QCD penguins of type two,
we appeal to flavor blindness to consider the set of four operators, 
\begin{eqnarray}
\D P^\prime_i
=
\D Q_i \Big|_{\psi_2 = \psi_3, \psi_4 = \psi_2}
\label{eq:pairs}
,\end{eqnarray}
that, 
by design, 
has only the requisite penguin contractions of type two. 
The mixing of 
$\D P^\prime_i$ 
operators with the penguin operators, 
$\D \cP_i$,
in Eq.~\eqref{eq:P2Chi}
is described by an anomalous dimension matrix
$\Gamma_{\D P^\prime}$, 
which is our desired matrix, 
$P^\prime = \Gamma_{\D P^\prime}$.

The use of Fierz identities is valuable here. 
In 't Hooft--Veltman dimensional regularization, 
the two-loop anomalous dimensions are the same between Fierz transformed operators. 
The operators in 
Eq.~\eqref{eq:pairs} 
are related by a Fierz transformation to operators with only type-one penguins. 
In essence, 
we are using the reverse of the argument employed by%
~\cite{Buras:1991jm} 
to determine penguin contributions.
The Fierz transformation of the 
$V_L \otimes V_L - V_R \otimes V_R$
operators produces two operators that we explicitly 
considered above in dealing with type-one penguins, 
namely
\begin{eqnarray}
\D P^\prime_1 
&=&
\D P_2, 
\notag \\
\D P^\prime_2
&=&
\D P_1
.\end{eqnarray}
On the other hand, 
the Fierz transformation of the two 
$V_L \otimes V_R - V_R \otimes V_L$
operators are themselves new operators,
\begin{eqnarray}
\D P^\prime_3 
&=&
2 \, ( \ol \psi_1 \psi_2]_L \, [\, \ol \psi_3 \psi_3 )_R
- 
\{ L \leftrightarrow R \}
,\notag \\
\D P^\prime_4
&=&
2 \, (\ol \psi_1 \psi_2)_L (\ol \psi_3 \psi_3)_R
-
\{ L \leftrightarrow R \}
\label{eq:NoPeng}
,\end{eqnarray}  
that involve only scalar bilinears. 
Because we utilize a mass-independent renormalization scheme, 
closed quark loops arising from scalar bilinears will always involve an odd number of 
Dirac matrices, 
and thereby vanish. 
Thus the lower two rows of 
$P^\prime$ 
must be zero.

From the results quoted above for 
$\D P_{1,2}$, 
we can easily determine the type-two penguin anomalous dimension matrix. 
For completeness, 
we display these matrices explicitly. 
The leading-order matrix has the form
\begin{equation}
(\gamma_{\D P^\prime})_0
=
\begin{pmatrix}
- \frac{2}{9} & \frac{2}{3} & - \frac{2}{9} & \frac{2}{3} 
\\ 
\phantom{-}0 & 0 & \phantom{-}0 & 0 
\\ 
\phantom{-}0 & 0 & \phantom{-}0 & 0 
\\
\phantom{-}0 & 0 & \phantom{-}0 & 0 
\end{pmatrix}
.\end{equation}
While at next-to-leading order, 
we have the matrix
\begin{equation}
(\gamma_{\D P^\prime})_1
= 
\begin{pmatrix}
- \frac{418}{243}
& 
\frac{850}{81}
& 
-\frac{1210}{243}
& 
\frac{490}{81}
\\
\\ 
\phantom{-}
\frac{23}{3}
& 
1 
& 
- 
\frac{25}{3}
& 
1 
\\
\\
\phantom{-}0
& 
0
& 
\phantom{-}0
& 
0
\\ \\
\phantom{-}0
& 
0
& 
\phantom{-}0
& 
0
\\
\end{pmatrix}
.\end{equation}

\subsubsection{Consistency Checks}

As a consistency check, 
one can easily verify that after a linear transformation, 
the matrices
$\cC_0$, 
$P_0$, 
and
$P^\prime_0$
lead to the same one-loop anomalous dimension matrix
$\gamma_0^{(N_f)}$
employed by~\cite{Dai:1991bx}.

A further consistency check is provided by the redundancy of the operator
$\mathcal{O}_4^{(2)}$ 
expressed by the Fierz relation in Eq.~\eqref{eq:FC}. 
To expose this redundancy, 
we can transform to a new basis specified by
$\mathcal{O}^{\prime \, (3)}_i = \sum_j \mathcal{S}_{ij} \, \mathcal{O}_j^{(3)}$, 
with the transformation matrix given by 
\begin{equation}
\mathcal{S}
=
\begin{pmatrix}
1 & \phantom{-} 0 & 0 & 0  &  \phantom{-} 0 & 0 \\
1 & -1 & 0 & 0 & -1 & 1  \\
0 & \phantom{-}0 & 1 & 0  & \phantom{-} 0 & 0 \\
0 & \phantom{-}0 & 0 & 1  & \phantom{-} 0 & 0 \\
0 & \phantom{-}0 & 0 & 0  & \phantom{-} 1 & 0 \\
0 & \phantom{-}0 & 0 & 0  & \phantom{-} 0 & 1 \\
\end{pmatrix}
.\end{equation}
Thus the primed basis is identical to the unprimed basis except for the redundant operator
$\mathcal{O}^{(3)}_2$. 
In the primed basis, 
the corresponding operator is
\begin{equation}
\mathcal{O}_2^{\prime \, (3)} = \mathcal{O}_1^{(3)} - \mathcal{O}_2^{(3)} - \mathcal{O}_5^{(3)} + \mathcal{O}_6^{(3)}
.\end{equation} 
The Fierz identity implies 
$\mathcal{O}_2^{\prime \, (3)} \equiv 0$, 
and accordingly this operator should not mix with any other operators. 
Remember that the 't Hooft--Veltman scheme preserves Fierz relations.

In terms of the anomalous dimension matrix in the primed basis, 
$\Gamma^{\prime \, (3)}$,
we must have the vanishing of the components
$\Gamma^{\prime \, (3)} _{2i} = 0$,
for any 
$i \neq 2$. 
This condition ensures that whatever the value of 
$C^{\prime  \, (3)}_2 (\mu)$, 
it will have no effect on the evolution of the other Wilson coefficients, 
$C^{\prime \, (3)}_{i \neq 2}(\mu)$, 
see Eq.~\eqref{eq:CWrun}.%
\footnote{
While the Wilson coefficient
$C_2^{\prime \, (3)}$
evolves with the renormalization scale 
according to
$\mu \frac{d}{d\mu} C_2^{\prime \, (3)} = \sum_{i} C_i^{\prime \, (3)} \Gamma^{\prime \, (3)}_{i2}  \neq 0$, 
the product 
$C_2^{\prime \, (3)} (\mu) \mathcal{O}_2^{\prime \, (3)} (\mu)$ 
is trivially scale independent because the Fierz identity implies
$\mathcal{O}_2^{ \prime\, (3)} (\mu) \equiv 0$, 
at all scales. 
Consequently any matrix elements of this operator computed by lattice techniques should vanish up to statistical noise. 
} 
The anomalous dimension matrix in the primed basis
is given by
$\Gamma^{\prime\, (3)} = \mathcal{S} \, \Gamma^{(3)} \mathcal{S}^{-1}$, 
where it happens that
$\mathcal{S}^{-1} = \mathcal{S}$. 
We have the constraint
\begin{eqnarray}
\Big[
\mathcal{S} \, \Gamma^{(3)} \mathcal{S}
\Big]_{2i}
= 0, \text{ for } i \neq 2
,\label{eq:Constr}
\end{eqnarray}
which must be satisfied order-by-order in perturbation theory, 
and the leading-order and next-to-leading order anomalous dimension
matrices indeed satisfy Eq.~\eqref{eq:Constr}.

To next-to-leading order, 
we actually find that
current-current contributions to 
Eq.~\eqref{eq:Constr} 
vanish independent of the value of
$N_f$
(recall that 
$\cC_1$
depends on 
$N_f$).
The remaining contributions to the constraint
involve the anomalous dimension matrices 
$P$ 
and 
$P^\prime$
with particular numerical coefficients depending on 
$N_f$. 
For these terms, 
only the contributions for the value
$N_f = 3$
vanish, 
and this is precisely the situation where the Fierz constraint exists.

\subsection{Results} \label{s:Results}%

With the anomalous dimension matrices fully specified, 
we can now run the Wilson coefficients down to hadronic scales.
We carry out the running in the chiral basis; 
and, 
for the reasons described above in Sec.~\ref{s:C}, 
we find is practical to convert to the 
$| \D I | = 1$
parity-violating operator basis given in~\cite{Kaplan:1992vj}.
This basis is specified by eight operators. 
There are four non-strange operators 
\begin{eqnarray}
O_1
&=&
( \ol u u - \ol d d )_A (\ol u u + \ol d d )_V,
\notag \\
O_2
&=&
( \ol u u - \ol d d \, ]_A  [ \, \ol u u + \ol d d )_V,
\notag \\
O_3
&=&
( \ol u u - \ol d d )_V  ( \ol u u + \ol d d  )_A,
\notag \\
O_4
&=&
( \ol u u - \ol d d \, ]_V  [ \, \ol u u + \ol d d )_A,
\label{eq:3PVKS}
\end{eqnarray}
and four strange operators
\begin{eqnarray}
O_5
&=&
( \ol u u - \ol d d )_A (\ol s s )_V,
\notag \\
O_6
&=&
( \ol u u - \ol d d \, ]_A  [ \, \ol s s  )_V,
\notag \\
O_7
&=&
( \ol u u - \ol d d )_V  ( \ol s s  )_A,
\notag \\
O_8
&=&
( \ol u u - \ol d d \, ]_V  [ \,  \ol s s )_A
\label{eq:KS}
.\end{eqnarray}

To convert between bases, 
it is convenient to append two operators to 
Eq.~\eqref{eq:SPV}, 
namely
\begin{eqnarray}
\cO_7^{(3)}
&=&
(\ol u \gamma_\mu u - \ol d \gamma_\mu d)_L
(\ol s \gamma^\mu s)_R
- 
\{ L \leftrightarrow R \}
\notag \\
\cO_8^{(3)}
&=&
(\ol u \gamma_\mu u - \ol d \gamma_\mu d ]_L
[ \, \ol s \gamma^\mu s)_R
- 
\{ L \leftrightarrow R \} 
.\end{eqnarray}
These operators necessarily have zero Wilson coefficients.
The set of operators 
$O_i$
is related to the set of eight operators
$\cO_i^{(3)}$
through a series of linear transformations, 
\begin{equation}
O_i = \sum_{jk} \cJ_{ij} \cT_{jk} \cO_k^{(3)}
,\end{equation} 
with the transformation matrix
$\mathcal{T}$
responsible for converting the chiral basis to the vector--axial-vector basis. 
It is specified by
$\cT = \diag ( T, T)$, 
with the 
$4 \times 4$
transformation matrix
$T$
given by
\begin{equation}
T
=
-
\begin{pmatrix}
1 &  \phantom{-} 1 \\
1 & - 1
\end{pmatrix}
\label{eq:T}
,\end{equation}
in  
$2 \times 2$ 
block form. 
The transformation matrix
$\cJ$
alters the flavor structure of the operators so that we arrive at a separation between 
strange and non-strange operators. 
Written as a  
$4 \times 4$ 
block matrix, 
we have
\begin{equation}
\cJ
= 
\begin{pmatrix}
1 & - 1 \\
0 & \phantom{-} 1 
\end{pmatrix}
\label{eq:J}
.\end{equation}
Finally, 
for the eight Wilson coefficients in the new basis,
we write  
$\mathfrak{C}_i (\mu)$,
and these are given by
\begin{equation}
\mathfrak{C}_i (\mu) = \sum_{jk} C_k^{(3)} (\mu) \cT^{-1}_{kj} \cJ^{-1}_{ki}
,\end{equation} 
with 
$C_{7,8}^{(3)} (\mu) \equiv 0$.

Notice that in the basis employed by~\cite{Kaplan:1992vj}, 
the Fierz transformation leads to the constraint
\begin{equation}
O_4
= 
O_1
- 
O_2
+ 
O_3
\label{eq:FCKS}
.\end{equation}
Finally as a consequence of the renormalization in the chiral basis, 
the operators 
\begin{eqnarray}
O^\prime_1
&=&
- O_1 + O_3 + O_5 - O_7
\notag \\
O^\prime_2
&=&
- O_2 + O_4 + O_6 - O_8
,\end{eqnarray}
have vanishing Wilson coefficients because they are not generated by QCD radiative corrections.

Computed values for the Wilson coefficients 
$\mathfrak{C}_i (\mu)$
at a scale of
$\mu = 1 \, \texttt{GeV}$
are collected in Table~\ref{t:results}. 
As inputs to this calculation, 
we use the masses~\cite{Nakamura:2010zzi}, 
\begin{align}
&
M_Z 
= 
91.2 \, \texttt{GeV},
&
M_W 
= 
80.4 \, \texttt{GeV}, 
\notag \\ 
&m_b
= 
4.19 \, \texttt{GeV}, 
&
m_c 
=
1.29 \, \texttt{GeV}
,\end{align}
where the latter two are 
$\ol {\text{MS}}$
masses,
and correspond to the masses in the subtraction scheme we employ. 
The value of the strong coupling is taken at the weak scale, 
\begin{equation}
\alpha_s (M_Z) = 0.118
.\end{equation}
From the two-loop expression for 
$\alpha_s(\mu)$
given in
Eq.~\eqref{eq:Alpha},
we find the value of the five-flavor QCD scale parameter
$\Lambda^{(5)} = 0.231 \, \texttt{GeV}$.
The requirement of continuity at 
$m_b$ 
and 
$m_c$
leads to the values
$\Lambda^{(4)} = 0.330 \, \texttt{GeV}$
and
$\Lambda^{(3)} = 0.376 \, \texttt{GeV}$, 
respectively.

\begin{table}
\caption{\label{t:results}
Values for the Wilson coefficients of the 
$| \Delta I | = 1$ 
parity-violating operators, 
$\mathfrak{C}_i (\mu)$, 
at hadronic scales.
We compare values obtained from leading-order 
(LO) 
evolution quoted in~\cite{Kaplan:1992vj} 
to results obtained in this work. 
In all cases, 
we normalize the coefficients by dividing by the tree-level coefficient
$|(c_0)_1| = \frac{1}{3} \sin^2 \theta_W$,
and the values quoted are at the scale
$\mu = 1 \, \texttt{GeV}$. 
Ingredients of our 
LO and next-to-leading order 
(NLO)
computations are described in the text. 
To make a meaningful comparison of the results, 
we eliminate the redundant operator 
$O_4$
using Eq.~\eqref{eq:FCKS}. 
}
\begin{center}
\begin{tabular}{|c|cccc|}
\hline 
\hline
& \multicolumn{4}{c|}{$\mathfrak{C}_i(1 \, \texttt{GeV}) \,  / \, |(c_0)_1|$} 
\tabularnewline
$\quad i \quad$
& 
$\,$ LO~\cite{Kaplan:1992vj} $$
& 
$\quad$ LO $\quad$
& 
$$ NLO (Z) 
&
$$ NLO (Z+W) 
\tabularnewline
\hline
\hline
1
&
\phantom{-}0.403
&
\phantom{-}0.264
&
-0.054
&
-0.055
\tabularnewline
2
&
\phantom{-}0.765
&
\phantom{-}0.981
&
\phantom{-}0.803
&
\phantom{-}0.810
\tabularnewline
3
&
-0.463
&
-0.592
&
-0.629
&
-0.627
\tabularnewline
4
&
$\phantom{-}0$
&
$\phantom{-}0$
&
$\phantom{-}0$
&
$\phantom{-}0$
\tabularnewline
\hline
5
&
\phantom{-}5.61
&
\phantom{-}5.97
&
\phantom{-}4.85
&
\phantom{-}5.09
\tabularnewline
6
&
-1.90
&
-2.30
&
-2.14
&
-2.55
\tabularnewline
7
&
\phantom{-}4.74
&
\phantom{-}5.12
&
\phantom{-}4.27
&
\phantom{-}4.51
\tabularnewline
8
&
-2.67
&
-3.29
&
-2.94
&
-3.36
\tabularnewline
\hline
\hline
\end{tabular} 
\end{center} 
\end{table}

The results shown in the table compare our leading and next-to-leading order computations of the renormalization of isovector parity-violating operators. 
Ingredients of these computation are summarized as follows:
\begin{enumerate}
\item 
Leading order (LO):
coefficients evolved with one-loop anomalous dimensions but with two-loop running coupling, 
tree-level matching at the weak scale and at heavy quark thresholds, 
charged-current interactions excluded.

\item
Next-to-leading order Z [NLO (Z)]:
coefficients evolved with two-loop anomalous dimensions along with the two-loop running coupling, 
next-to-leading order matching at the weak scale and at heavy quark thresholds,
charged-current interactions excluded. 

\item
Next-to-leading order Z+W [NLO (Z+W)]:
same as NLO (Z), 
but now including charged-current interactions at tree level. 

\end{enumerate}

\noindent
We also compare our values with the leading-order evolution of Wilson coefficients
determined in~\cite{Dai:1991bx}, 
and quoted at 
$\mu = 1 \, \texttt{GeV}$ 
in~\cite{Kaplan:1992vj}. 
To present these and our results, 
moreover,
we do not give values for all eight coefficients. 
The Fierz constraint on the operators, 
Eq.~\eqref{eq:FCKS},
makes reporting values for the first four coefficients somewhat arbitrary. 
To eliminate this ambiguity, 
we choose 
$O_4$
as the redundant operator, 
and accordingly determine contributions to 
$\mathfrak{C}_{1\text{--}3}(\mu)$
from its coefficient
$\mathfrak{C}_4(\mu)$.

In the table, 
we see that from comparing leading and next-to-leading order results, 
the corrections are generally on the order of
$10$--$20 \%$. 
This is consistent with the typical size of next-to-leading order effects in the context of the
QCD evolution of flavor-changing weak currents. 
We also see that inclusion of isovector charged-current interactions has little effect on the 
coefficients of non-strange operators:
the first three coefficients are altered only by a few percent when the charged-current contributions are turned on. 
This behavior is not surprising given that the charged-current operators only mix with 
non-strange operators through penguin diagrams,
and this happens in the QCD evolution, 
moreover, 
only when 
$\mu < m_c$. 
Were we to evolve the coefficients down to lower scales, 
charged-current interactions would have a larger effect on the non-strange operator coefficients. 
The effect of charged-current interactions on the coefficients of strange operators, 
by contrast, 
is at the $10$--$20 \%$ level.

Finally,
the behavior of the coefficient
$\mathfrak{C}_1(\mu)$
gives one reason to pause. 
Our value at leading order is quite different from that in~\cite{Kaplan:1992vj}, 
which is already an indication of its sensitivity to higher-order corrections. 
The difference between leading and next-to-leading order is 
$120 \%$ 
of the leading-order value. 
While the size of such corrections is alarming, 
this coefficient is the smallest 
(in absolute value) 
of all hadronic-scale coefficients. 
This is true even if we use the leading-order value for 
$\mathfrak{C}_1(\mu)$.
Assuming the isovector parity-violating hadronic matrix elements of \emph{all}  
$O_i(\mu)$
operators are roughly the same size, 
we should not compare the difference of leading and next-to-leading order 
contributions to the value of 
$\mathfrak{C}_1(\mu)$, 
\emph{per se}, 
but rather
to the maximum value of all the 
$\mathfrak{C}_i(\mu)$
coefficients. 
The next-to-leading order corrections to 
$\mathfrak{C}_1(\mu)$
are only 
$6 \%$
of the maximum Wilson coefficient. 
Overall the next-to-leading order corrections to isovector parity-violating hadronic couplings should be under control.
This can be spoiled in practice if parity-violating hadronic matrix elements of 
$O_1(\mu)$
are dynamically enhanced an order of magnitude over the other operators, 
or if there are dramatic cancelations among the other terms.

\section{Outlook} \label{s:LookOut}%

Above, 
we study the QCD evolution of isovector parity-violating four-quark operators from the 
weak scale down to hadronic scales. 
At hadronic scales, 
matrix elements of the corresponding operators are required to determine the 
$| \Delta I | = 1$
parity-violating couplings 
between hadrons non-perturbatively using lattice QCD techniques.  
We find that next-to-leading order contributions alter the values of Wilson coefficients at 
$\mu = 1 \, \texttt{GeV}$
by 
$\sim 10 - 20 \%$
in most cases.
To compute these values, 
we determine the one-loop matching of Wilson coefficients at the weak scale, 
Eq.~\eqref{eq:ZMatch}, 
and deduce the form of isovector parity violation from the charged-current interaction, 
Eq.~\eqref{eq:WMatch}.
Appealing to parity invariance and the flavor-blindness of QCD radiation, 
we extract the required two-loop anomalous dimension matrices from those arising 
in the study of QCD corrections to 
$|\Delta S| =1$  
non-leptonic weak decays~\cite{Buras:1992tc}. 
Results are given in the dimensional regularization scheme of 't Hooft--Veltman, 
with an additional multiplicative renormalization of the axial-vector current
which is necessary in order to preserve chirality. 
In this scheme, 
we find that next-to-leading order corrections to Wilson coefficients are important. 
Moreover, 
inclusion of 
$| \Delta I | = 1$
charged-current interactions affects the coefficients of strange operators at hadronic scales by 
$10$--$20 \%$.

Having determined the Wilson coefficients of isovector parity-violating operators at hadronic scales
to next-to-leading order accuracy, 
there are a number of further studies that are now open to pursuit. 
First is the inclusion of electromagnetic corrections. 
In this work, 
we focus solely on QCD corrections; 
and,
at hadronic scales, 
$\mu$,
we include corrections that are of the order
$[\a_s(\mu) / (4 \pi)]^2$. 
The leading electromagnetic corrections arise proportional to  
$\a / 4 \pi$,
and are thus expected to be of the same size. 
Further study is required to determine the effects of electromagnetism on parity-violating operators at hadronic scales, 
and this involves mixing between the various isospin channels. 
A second investigation is the study QCD renormalization of parity violation using a  lattice regularization. 
One-loop computations using lattice perturbation theory are required to determine the renormalization scheme dependence, 
and knowledge of this result will enable one to convert a lattice regularization to the 
$\ol{\text{MS}}$
scheme. 
Such knowledge is essential to make physical predictions from lattice QCD computations of hadronic parity violation. 
To this end, 
a final investigation is to study the evolution of Wilson coefficients using a non-perturbative renormalization scheme, 
such as the renormalization independent---momentum scheme~\cite{Martinelli:1994ty}. 
Nevertheless, 
the investigation of hadronic parity violation represents an exciting opportunity for lattice QCD. 
In light of major experimental efforts, 
the lattice approach will be complimentary in mapping out the parity-violating nuclear force, 
and will ultimately connect these few-body parameters to the hadronic weak interaction in the Standard Model.

\begin{acknowledgments}
Work supported in part by a joint CCNY--RBRC fellowship, 
and by the Research Foundation of the CUNY.
We thank J.~Wasem for enlightening discussions, 
and the hospitality of Lawrence Livermore and Lawrence Berkeley National Laboratories during visits in the course of this work. 
\end{acknowledgments}

\appendix

\bibliography{hb}

\begin{thebibliography}{10}%
\makeatletter
\providecommand \@ifxundefined [1]{%
 \ifx #1\undefined \expandafter \@firstoftwo
 \else \expandafter \@secondoftwo
\fi
}%
\providecommand \@ifnum [1]{%
 \ifnum #1\expandafter \@firstoftwo
 \else \expandafter \@secondoftwo
\fi
}%
\providecommand \enquote [1]{``#1''}%
\providecommand \bibnamefont  [1]{#1}%
\providecommand \bibfnamefont [1]{#1}%
\providecommand \citenamefont [1]{#1}%
\providecommand\href[0]{\@sanitize\@href}%
\providecommand\@href[1]{\endgroup\@@startlink{#1}\endgroup\@@href}%
\providecommand\@@href[1]{#1\@@endlink}%
\providecommand \@sanitize [0]{\begingroup\catcode`\&12\catcode`\#12\relax}%
\@ifxundefined \pdfoutput {\@firstoftwo}{%
 \@ifnum{\z@=\pdfoutput}{\@firstoftwo}{\@secondoftwo}%
}{%
 \providecommand\@@startlink[1]{\leavevmode\special{html:<a href="#1">}}%
 \providecommand\@@endlink[0]{\special{html:</a>}}%
}{%
 \providecommand\@@startlink[1]{%
  \leavevmode
  \pdfstartlink
   attr{/Border[0 0 1 ]/H/I/C[0 1 1]}%
   user{/Subtype/Link/A<</Type/Action/S/URI/URI(#1)>>}%
  \relax
 }%
 \providecommand\@@endlink[0]{\pdfendlink}%
}%
\providecommand \url  [0]{\begingroup\@sanitize \@url }%
\providecommand \@url [1]{\endgroup\@href {#1}{\urlprefix}}%
\providecommand \urlprefix [0]{URL }%
\providecommand \Eprint[0]{\href }%
\@ifxundefined \urlstyle {%
  \providecommand \doi [1]{doi:\discretionary{}{}{}#1}%
}{%
  \providecommand \doi [0]{doi:\discretionary{}{}{}\begingroup
  \urlstyle{rm}\Url }%
}%
\providecommand \doibase [0]{http://dx.doi.org/}%
\providecommand \Doi[1]{\href{\doibase#1}}%
\providecommand \bibAnnote [3]{%
  \BibitemShut{#1}%
  \begin{quotation}\noindent
    \textsc{Key:}\ #2\\\textsc{Annotation:}\ #3%
  \end{quotation}%
}%
\providecommand \bibAnnoteFile [2]{%
  \IfFileExists{#2}{\bibAnnote {#1} {#2} {\input{#2}}}{}%
}%
\providecommand \typeout [0]{\immediate \write \m@ne }%
\providecommand \selectlanguage [0]{\@gobble}%
\providecommand \bibinfo [0]{\@secondoftwo}%
\providecommand \bibfield [0]{\@secondoftwo}%
\providecommand \translation [1]{[#1]}%
\providecommand \BibitemOpen[0]{}%
\providecommand \bibitemStop [0]{}%
\providecommand \bibitemNoStop [0]{.\EOS\space}%
\providecommand \EOS [0]{\spacefactor3000\relax}%
\providecommand \BibitemShut [1]{\csname bibitem#1\endcsname}%
\bibitem{Glashow:1970gm}%
  \BibitemOpen
  \bibfield{author}{%
  \bibinfo {author} {\bibfnamefont{S.}~\bibnamefont{Glashow}}, \bibinfo
  {author} {\bibfnamefont{J.}~\bibnamefont{Iliopoulos}},\ and\ \bibinfo
  {author} {\bibfnamefont{L.}~\bibnamefont{Maiani}},\ }%
  \bibfield{journal}{%
  \Doi{10.1103/PhysRevD.2.1285}{\bibinfo {journal} {Phys.Rev.}}\ }%
  \textbf{\bibinfo {volume} {D2}},\ \bibinfo {pages} {1285} (\bibinfo {year}
  {1970})%
  \bibAnnoteFile{NoStop}{Glashow:1970gm}%
\bibitem{Tanner:1957zz}%
  \BibitemOpen
  \bibfield{author}{%
  \bibinfo {author} {\bibfnamefont{N.}~\bibnamefont{Tanner}},\ }%
  \bibfield{journal}{%
  \Doi{10.1103/PhysRev.107.1203}{\bibinfo {journal} {Phys.Rev.}}\ }%
  \textbf{\bibinfo {volume} {107}},\ \bibinfo {pages} {1203} (\bibinfo {year}
  {1957})%
  \bibAnnoteFile{NoStop}{Tanner:1957zz}%
\bibitem{Lobashov:1967aa}%
  \BibitemOpen
  \bibfield{author}{%
  \bibinfo {author} {\bibfnamefont{V.}~\bibnamefont{Lobashov}}, \bibinfo
  {author} {\bibfnamefont{V.}~\bibnamefont{Nazarenko}}, \bibinfo {author}
  {\bibfnamefont{L.}~\bibnamefont{Saenko}}, \bibinfo {author}
  {\bibfnamefont{L.}~\bibnamefont{Smotritsky}},\ and\ \bibinfo {author}
  {\bibfnamefont{G.}~\bibnamefont{Kharkevitch}},\ }%
  \bibfield{journal}{%
  \Doi{10.1016/0370-2693(67)90191-8}{\bibinfo {journal} {Phys.Lett.}}\ }%
  \textbf{\bibinfo {volume} {B25}},\ \bibinfo {pages} {104} (\bibinfo {year}
  {1967})%
  \bibAnnoteFile{NoStop}{Lobashov:1967aa}%
\bibitem{Bowman:1989ci}%
  \BibitemOpen
  \bibfield{author}{%
  \bibinfo {author} {\bibfnamefont{C.}~\bibnamefont{Bowman}}, \bibinfo {author}
  {\bibfnamefont{J.}~\bibnamefont{Bowman}},\ and\ \bibinfo {author}
  {\bibfnamefont{V.}~\bibnamefont{Yuan}},\ }%
  \bibfield{journal}{%
  \Doi{10.1103/PhysRevC.39.1721}{\bibinfo {journal} {Phys.Rev.}}\ }%
  \textbf{\bibinfo {volume} {C39}},\ \bibinfo {pages} {1721} (\bibinfo {year}
  {1989})%
  \bibAnnoteFile{NoStop}{Bowman:1989ci}%
\bibitem{Adelberger:1985ik}%
  \BibitemOpen
  \bibfield{author}{%
  \bibinfo {author} {\bibfnamefont{E.}~\bibnamefont{Adelberger}}\ and\ \bibinfo
  {author} {\bibfnamefont{W.}~\bibnamefont{Haxton}},\ }%
  \bibfield{journal}{%
  \bibinfo {journal} {Ann.~Rev.~Nucl.~Part.~Sci.}\ }%
  \textbf{\bibinfo {volume} {35}},\ \bibinfo {pages} {501} (\bibinfo {year}
  {1985})%
  \bibAnnoteFile{NoStop}{Adelberger:1985ik}%
\bibitem{RamseyMusolf:2006dz}%
  \BibitemOpen
  \bibfield{author}{%
  \bibinfo {author} {\bibfnamefont{M.~J.}\ \bibnamefont{Ramsey-Musolf}}\ and\
  \bibinfo {author} {\bibfnamefont{S.~A.}\ \bibnamefont{Page}},\ }%
  \bibfield{journal}{%
  \Doi{10.1146/annurev.nucl.54.070103.181255}{\bibinfo {journal}
  {Ann.~Rev.~Nucl.~Part.~Sci.}}\ }%
  \textbf{\bibinfo {volume} {56}},\ \bibinfo {pages} {1} (\bibinfo {year}
  {2006}),\ \Eprint{http://arxiv.org/abs/hep-ph/0601127}{arXiv:hep-ph/0601127
  [hep-ph]}%
  \bibAnnoteFile{NoStop}{RamseyMusolf:2006dz}%
\bibitem{Snow:2011zz}%
  \BibitemOpen
  \bibfield{author}{%
  \bibinfo {author} {\bibfnamefont{W.~M.}\ \bibnamefont{Snow}} \emph{et~al.},\
  }%
  \bibfield{journal}{%
  \Doi{10.1103/PhysRevC.83.022501}{\bibinfo {journal} {Phys.Rev.}}\ }%
  \textbf{\bibinfo {volume} {C83}},\ \bibinfo {pages} {022501} (\bibinfo {year}
  {2011})%
  \bibAnnoteFile{NoStop}{Snow:2011zz}%
\bibitem{Gericke:2011zz}%
  \BibitemOpen
  \bibfield{author}{%
  \bibinfo {author} {\bibfnamefont{M.~T.}\ \bibnamefont{Gericke}}
  \emph{et~al.},\ }%
  \bibfield{journal}{%
  \Doi{10.1103/PhysRevC.83.015505}{\bibinfo {journal} {Phys.Rev.}}\ }%
  \textbf{\bibinfo {volume} {C83}},\ \bibinfo {pages} {015505} (\bibinfo {year}
  {2011})%
  \bibAnnoteFile{NoStop}{Gericke:2011zz}%
\bibitem{Desplanques:1979hn}%
  \BibitemOpen
  \bibfield{author}{%
  \bibinfo {author} {\bibfnamefont{B.}~\bibnamefont{Desplanques}}, \bibinfo
  {author} {\bibfnamefont{J.~F.}\ \bibnamefont{Donoghue}},\ and\ \bibinfo
  {author} {\bibfnamefont{B.~R.}\ \bibnamefont{Holstein}},\ }%
  \bibfield{journal}{%
  \Doi{10.1016/0003-4916(80)90217-1}{\bibinfo {journal} {Annals Phys.}}\ }%
  \textbf{\bibinfo {volume} {124}},\ \bibinfo {pages} {449} (\bibinfo {year}
  {1980})%
  \bibAnnoteFile{NoStop}{Desplanques:1979hn}%
\bibitem{Zhu:2004vw}%
  \BibitemOpen
  \bibfield{author}{%
  \bibinfo {author} {\bibfnamefont{S.-L.}\ \bibnamefont{Zhu}}, \bibinfo
  {author} {\bibfnamefont{C.}~\bibnamefont{Maekawa}}, \bibinfo {author}
  {\bibfnamefont{B.}~\bibnamefont{Holstein}}, \bibinfo {author}
  {\bibfnamefont{M.}~\bibnamefont{Ramsey-Musolf}},\ and\ \bibinfo {author}
  {\bibfnamefont{U.}~\bibnamefont{van Kolck}},\ }%
  \bibfield{journal}{%
  \Doi{10.1016/j.nuclphysa.2004.10.032}{\bibinfo {journal} {Nucl.Phys.}}\ }%
  \textbf{\bibinfo {volume} {A748}},\ \bibinfo {pages} {435} (\bibinfo {year}
  {2005}),\ \Eprint{http://arxiv.org/abs/nucl-th/0407087}{arXiv:nucl-th/0407087
  [nucl-th]}%
  \bibAnnoteFile{NoStop}{Zhu:2004vw}%
\bibitem{Phillips:2008hn}%
  \BibitemOpen
  \bibfield{author}{%
  \bibinfo {author} {\bibfnamefont{D.~R.}\ \bibnamefont{Phillips}}, \bibinfo
  {author} {\bibfnamefont{M.~R.}\ \bibnamefont{Schindler}},\ and\ \bibinfo
  {author} {\bibfnamefont{R.~P.}\ \bibnamefont{Springer}},\ }%
  \bibfield{journal}{%
  \Doi{10.1016/j.nuclphysa.2009.02.011}{\bibinfo {journal} {Nucl.Phys.}}\ }%
  \textbf{\bibinfo {volume} {A822}},\ \bibinfo {pages} {1} (\bibinfo {year}
  {2009}),\ \Eprint{http://arxiv.org/abs/0812.2073}{arXiv:0812.2073 [nucl-th]}%
  \bibAnnoteFile{NoStop}{Phillips:2008hn}%
\bibitem{Bedaque:2002mn}%
  \BibitemOpen
  \bibfield{author}{%
  \bibinfo {author} {\bibfnamefont{P.~F.}\ \bibnamefont{Bedaque}}\ and\
  \bibinfo {author} {\bibfnamefont{U.}~\bibnamefont{van Kolck}},\ }%
  \bibfield{journal}{%
  \Doi{10.1146/annurev.nucl.52.050102.090637}{\bibinfo {journal}
  {Ann.Rev.Nucl.Part.Sci.}}\ }%
  \textbf{\bibinfo {volume} {52}},\ \bibinfo {pages} {339} (\bibinfo {year}
  {2002}),\ \Eprint{http://arxiv.org/abs/nucl-th/0203055}{arXiv:nucl-th/0203055
  [nucl-th]}%
  \bibAnnoteFile{NoStop}{Bedaque:2002mn}%
\bibitem{Girlanda:2008ts}%
  \BibitemOpen
  \bibfield{author}{%
  \bibinfo {author} {\bibfnamefont{L.}~\bibnamefont{Girlanda}},\ }%
  \bibfield{journal}{%
  \Doi{10.1103/PhysRevC.77.067001}{\bibinfo {journal} {Phys.Rev.}}\ }%
  \textbf{\bibinfo {volume} {C77}},\ \bibinfo {pages} {067001} (\bibinfo {year}
  {2008}),\ \Eprint{http://arxiv.org/abs/0804.0772}{arXiv:0804.0772 [nucl-th]}%
  \bibAnnoteFile{NoStop}{Girlanda:2008ts}%
\bibitem{Shin:2009hi}%
  \BibitemOpen
  \bibfield{author}{%
  \bibinfo {author} {\bibfnamefont{J.}~\bibnamefont{Shin}}, \bibinfo {author}
  {\bibfnamefont{S.}~\bibnamefont{Ando}},\ and\ \bibinfo {author}
  {\bibfnamefont{C.}~\bibnamefont{Hyun}},\ }%
  \bibfield{journal}{%
  \Doi{10.1103/PhysRevC.81.055501}{\bibinfo {journal} {Phys.Rev.}}\ }%
  \textbf{\bibinfo {volume} {C81}},\ \bibinfo {pages} {055501} (\bibinfo {year}
  {2010}),\ \Eprint{http://arxiv.org/abs/0907.3995}{arXiv:0907.3995 [nucl-th]}%
  \bibAnnoteFile{NoStop}{Shin:2009hi}%
\bibitem{Schindler:2009wd}%
  \BibitemOpen
  \bibfield{author}{%
  \bibinfo {author} {\bibfnamefont{M.~R.}\ \bibnamefont{Schindler}}\ and\
  \bibinfo {author} {\bibfnamefont{R.~P.}\ \bibnamefont{Springer}},\ }%
  \bibfield{journal}{%
  \Doi{10.1016/j.nuclphysa.2010.06.002}{\bibinfo {journal} {Nucl.Phys.}}\ }%
  \textbf{\bibinfo {volume} {A846}},\ \bibinfo {pages} {51} (\bibinfo {year}
  {2010}),\ \Eprint{http://arxiv.org/abs/0907.5358}{arXiv:0907.5358 [nucl-th]}%
  \bibAnnoteFile{NoStop}{Schindler:2009wd}%
\bibitem{Griesshammer:2011md}%
  \BibitemOpen
  \bibfield{author}{%
  \bibinfo {author} {\bibfnamefont{H.~W.}\ \bibnamefont{Griesshammer}},
  \bibinfo {author} {\bibfnamefont{M.~R.}\ \bibnamefont{Schindler}},\ and\
  \bibinfo {author} {\bibfnamefont{R.~P.}\ \bibnamefont{Springer}}}%
   (\bibinfo {year} {2011}),\
  \Eprint{http://arxiv.org/abs/1109.5667}{arXiv:1109.5667 [nucl-th]}%
  \bibAnnoteFile{NoStop}{Griesshammer:2011md}%
\bibitem{Vanasse:2011nd}%
  \BibitemOpen
  \bibfield{author}{%
  \bibinfo {author} {\bibfnamefont{J.}~\bibnamefont{Vanasse}}}%
   (\bibinfo {year} {2011}),\
  \Eprint{http://arxiv.org/abs/1110.1039}{arXiv:1110.1039 [nucl-th]}%
  \bibAnnoteFile{NoStop}{Vanasse:2011nd}%
\bibitem{Kaplan:1992vj}%
  \BibitemOpen
  \bibfield{author}{%
  \bibinfo {author} {\bibfnamefont{D.~B.}\ \bibnamefont{Kaplan}}\ and\ \bibinfo
  {author} {\bibfnamefont{M.~J.}\ \bibnamefont{Savage}},\ }%
  \bibfield{journal}{%
  \Doi{10.1016/0375-9474(93)90475-D, 10.1016/0375-9474(93)90475-D}{\bibinfo
  {journal} {Nucl.Phys.}}\ }%
  \textbf{\bibinfo {volume} {A556}},\ \bibinfo {pages} {653} (\bibinfo {year}
  {1993})%
  \bibAnnoteFile{NoStop}{Kaplan:1992vj}%
\bibitem{Zhu:2000fc}%
  \BibitemOpen
  \bibfield{author}{%
  \bibinfo {author} {\bibfnamefont{S.-L.}\ \bibnamefont{Zhu}}, \bibinfo
  {author} {\bibfnamefont{S.}~\bibnamefont{Puglia}}, \bibinfo {author}
  {\bibfnamefont{B.~R.}\ \bibnamefont{Holstein}},\ and\ \bibinfo {author}
  {\bibfnamefont{M.}~\bibnamefont{Ramsey-Musolf}},\ }%
  \bibfield{journal}{%
  \Doi{10.1103/PhysRevD.63.033006}{\bibinfo {journal} {Phys.Rev.}}\ }%
  \textbf{\bibinfo {volume} {D63}},\ \bibinfo {pages} {033006} (\bibinfo {year}
  {2001}),\ \Eprint{http://arxiv.org/abs/hep-ph/0005281}{arXiv:hep-ph/0005281
  [hep-ph]}%
  \bibAnnoteFile{NoStop}{Zhu:2000fc}%
\bibitem{Bedaque:1999dh}%
  \BibitemOpen
  \bibfield{author}{%
  \bibinfo {author} {\bibfnamefont{P.~F.}\ \bibnamefont{Bedaque}}\ and\
  \bibinfo {author} {\bibfnamefont{M.~J.}\ \bibnamefont{Savage}},\ }%
  \bibfield{journal}{%
  \Doi{10.1103/PhysRevC.62.018501}{\bibinfo {journal} {Phys.Rev.}}\ }%
  \textbf{\bibinfo {volume} {C62}},\ \bibinfo {pages} {018501} (\bibinfo {year}
  {2000}),\ \Eprint{http://arxiv.org/abs/nucl-th/9909055}{arXiv:nucl-th/9909055
  [nucl-th]}%
  \bibAnnoteFile{NoStop}{Bedaque:1999dh}%
\bibitem{Chen:2000mb}%
  \BibitemOpen
  \bibfield{author}{%
  \bibinfo {author} {\bibfnamefont{J.-W.}\ \bibnamefont{Chen}}, \bibinfo
  {author} {\bibfnamefont{T.~D.}\ \bibnamefont{Cohen}},\ and\ \bibinfo {author}
  {\bibfnamefont{C.-W.}\ \bibnamefont{Kao}},\ }%
  \bibfield{journal}{%
  \Doi{10.1103/PhysRevC.64.055206}{\bibinfo {journal} {Phys.Rev.}}\ }%
  \textbf{\bibinfo {volume} {C64}},\ \bibinfo {pages} {055206} (\bibinfo {year}
  {2001}),\ \Eprint{http://arxiv.org/abs/nucl-th/0009031}{arXiv:nucl-th/0009031
  [nucl-th]}%
  \bibAnnoteFile{NoStop}{Chen:2000mb}%
\bibitem{Chen:2000hb}%
  \BibitemOpen
  \bibfield{author}{%
  \bibinfo {author} {\bibfnamefont{J.-W.}\ \bibnamefont{Chen}}\ and\ \bibinfo
  {author} {\bibfnamefont{X.-D.}\ \bibnamefont{Ji}},\ }%
  \bibfield{journal}{%
  \Doi{10.1103/PhysRevLett.86.4239}{\bibinfo {journal} {Phys.Rev.Lett.}}\ }%
  \textbf{\bibinfo {volume} {86}},\ \bibinfo {pages} {4239} (\bibinfo {year}
  {2001}),\ \Eprint{http://arxiv.org/abs/hep-ph/0011230}{arXiv:hep-ph/0011230
  [hep-ph]}%
  \bibAnnoteFile{NoStop}{Chen:2000hb}%
\bibitem{Chen:2000km}%
  \BibitemOpen
  \bibfield{author}{%
  \bibinfo {author} {\bibfnamefont{J.-W.}\ \bibnamefont{Chen}}\ and\ \bibinfo
  {author} {\bibfnamefont{X.-D.}\ \bibnamefont{Ji}},\ }%
  \bibfield{journal}{%
  \Doi{10.1016/S0370-2693(01)00100-9}{\bibinfo {journal} {Phys.Lett.}}\ }%
  \textbf{\bibinfo {volume} {B501}},\ \bibinfo {pages} {209} (\bibinfo {year}
  {2001}),\ \Eprint{http://arxiv.org/abs/nucl-th/0011100}{arXiv:nucl-th/0011100
  [nucl-th]}%
  \bibAnnoteFile{NoStop}{Chen:2000km}%
\bibitem{Beane:2002ca}%
  \BibitemOpen
  \bibfield{author}{%
  \bibinfo {author} {\bibfnamefont{S.~R.}\ \bibnamefont{Beane}}\ and\ \bibinfo
  {author} {\bibfnamefont{M.~J.}\ \bibnamefont{Savage}},\ }%
  \bibfield{journal}{%
  \Doi{10.1016/S0550-3213(02)00405-4}{\bibinfo {journal} {Nucl.Phys.}}\ }%
  \textbf{\bibinfo {volume} {B636}},\ \bibinfo {pages} {291} (\bibinfo {year}
  {2002}),\ \Eprint{http://arxiv.org/abs/hep-lat/0203028}{arXiv:hep-lat/0203028
  [hep-lat]}%
  \bibAnnoteFile{NoStop}{Beane:2002ca}%
\bibitem{Tiburzi:2004mv}%
  \BibitemOpen
  \bibfield{author}{%
  \bibinfo {author} {\bibfnamefont{B.~C.}\ \bibnamefont{Tiburzi}},\ }%
  \bibfield{journal}{%
  \Doi{10.1103/PhysRevD.71.054504}{\bibinfo {journal} {Phys.Rev.}}\ }%
  \textbf{\bibinfo {volume} {D71}},\ \bibinfo {pages} {054504} (\bibinfo {year}
  {2005}),\ \Eprint{http://arxiv.org/abs/hep-lat/0412025}{arXiv:hep-lat/0412025
  [hep-lat]}%
  \bibAnnoteFile{NoStop}{Tiburzi:2004mv}%
\bibitem{Wasem:2011tp}%
  \BibitemOpen
  \bibfield{author}{%
  \bibinfo {author} {\bibfnamefont{J.}~\bibnamefont{Wasem}}}%
   (\bibinfo {year} {2011}),\
  \Eprint{http://arxiv.org/abs/1108.1151}{arXiv:1108.1151 [hep-lat]}%
  \bibAnnoteFile{NoStop}{Wasem:2011tp}%
\bibitem{Blum:2011pu}%
  \BibitemOpen
  \bibfield{author}{%
  \bibinfo {author} {\bibfnamefont{T.}~\bibnamefont{Blum}}, \bibinfo {author}
  {\bibfnamefont{P.}~\bibnamefont{Boyle}}, \bibinfo {author}
  {\bibfnamefont{N.}~\bibnamefont{Christ}}, \bibinfo {author}
  {\bibfnamefont{N.}~\bibnamefont{Garron}}, \bibinfo {author}
  {\bibfnamefont{E.}~\bibnamefont{Goode}}, \emph{et~al.},\ }%
  \bibfield{journal}{%
  \Doi{10.1103/PhysRevD.84.114503}{\bibinfo {journal} {Phys.Rev.}}\ }%
  \textbf{\bibinfo {volume} {D84}},\ \bibinfo {pages} {114503} (\bibinfo {year}
  {2011}),\ \Eprint{http://arxiv.org/abs/1106.2714}{arXiv:1106.2714 [hep-lat]}%
  \bibAnnoteFile{NoStop}{Blum:2011pu}%
\bibitem{Blum:2011ng}%
  \BibitemOpen
  \bibfield{author}{%
  \bibinfo {author} {\bibfnamefont{T.}~\bibnamefont{Blum}}, \bibinfo {author}
  {\bibfnamefont{P.}~\bibnamefont{Boyle}}, \bibinfo {author}
  {\bibfnamefont{N.}~\bibnamefont{Christ}}, \bibinfo {author}
  {\bibfnamefont{N.}~\bibnamefont{Garron}}, \bibinfo {author}
  {\bibfnamefont{E.}~\bibnamefont{Goode}}, \emph{et~al.}}%
   (\bibinfo {year} {2011}),\
  \Eprint{http://arxiv.org/abs/1111.1699}{arXiv:1111.1699 [hep-lat]}%
  \bibAnnoteFile{NoStop}{Blum:2011ng}%
\bibitem{Dai:1991bx}%
  \BibitemOpen
  \bibfield{author}{%
  \bibinfo {author} {\bibfnamefont{J.}~\bibnamefont{Dai}}, \bibinfo {author}
  {\bibfnamefont{M.~J.}\ \bibnamefont{Savage}}, \bibinfo {author}
  {\bibfnamefont{J.}~\bibnamefont{Liu}},\ and\ \bibinfo {author}
  {\bibfnamefont{R.~P.}\ \bibnamefont{Springer}},\ }%
  \bibfield{journal}{%
  \Doi{10.1016/0370-2693(91)90108-3}{\bibinfo {journal} {Phys.Lett.}}\ }%
  \textbf{\bibinfo {volume} {B271}},\ \bibinfo {pages} {403} (\bibinfo {year}
  {1991})%
  \bibAnnoteFile{NoStop}{Dai:1991bx}%
\bibitem{Buchalla:1995vs}%
  \BibitemOpen
  \bibfield{author}{%
  \bibinfo {author} {\bibfnamefont{G.}~\bibnamefont{Buchalla}}, \bibinfo
  {author} {\bibfnamefont{A.~J.}\ \bibnamefont{Buras}},\ and\ \bibinfo {author}
  {\bibfnamefont{M.~E.}\ \bibnamefont{Lautenbacher}},\ }%
  \bibfield{journal}{%
  \Doi{10.1103/RevModPhys.68.1125}{\bibinfo {journal} {Rev.Mod.Phys.}}\ }%
  \textbf{\bibinfo {volume} {68}},\ \bibinfo {pages} {1125} (\bibinfo {year}
  {1996}),\ \Eprint{http://arxiv.org/abs/hep-ph/9512380}{arXiv:hep-ph/9512380
  [hep-ph]}%
  \bibAnnoteFile{NoStop}{Buchalla:1995vs}%
\bibitem{'tHooft:1972fi}%
  \BibitemOpen
  \bibfield{author}{%
  \bibinfo {author} {\bibfnamefont{G.}~\bibnamefont{'t~Hooft}}\ and\ \bibinfo
  {author} {\bibfnamefont{M.}~\bibnamefont{Veltman}},\ }%
  \bibfield{journal}{%
  \Doi{10.1016/0550-3213(72)90279-9}{\bibinfo {journal} {Nucl.Phys.}}\ }%
  \textbf{\bibinfo {volume} {B44}},\ \bibinfo {pages} {189} (\bibinfo {year}
  {1972})%
  \bibAnnoteFile{NoStop}{'tHooft:1972fi}%
\bibitem{Akyeampong:1973xi}%
  \BibitemOpen
  \bibfield{author}{%
  \bibinfo {author} {\bibfnamefont{D.}~\bibnamefont{Akyeampong}}\ and\ \bibinfo
  {author} {\bibfnamefont{R.}~\bibnamefont{Delbourgo}},\ }%
  \bibfield{journal}{%
  \Doi{10.1007/BF02786835}{\bibinfo {journal} {Nuovo Cim.}}\ }%
  \textbf{\bibinfo {volume} {A17}},\ \bibinfo {pages} {578} (\bibinfo {year}
  {1973})%
  \bibAnnoteFile{NoStop}{Akyeampong:1973xi}%
\bibitem{Breitenlohner:1977hr}%
  \BibitemOpen
  \bibfield{author}{%
  \bibinfo {author} {\bibfnamefont{P.}~\bibnamefont{Breitenlohner}}\ and\
  \bibinfo {author} {\bibfnamefont{D.}~\bibnamefont{Maison}},\ }%
  \bibfield{journal}{%
  \Doi{10.1007/BF01609069}{\bibinfo {journal} {Commun.Math.Phys.}}\ }%
  \textbf{\bibinfo {volume} {52}},\ \bibinfo {pages} {11} (\bibinfo {year}
  {1977})%
  \bibAnnoteFile{NoStop}{Breitenlohner:1977hr}%
\bibitem{Buras:1992tc}%
  \BibitemOpen
  \bibfield{author}{%
  \bibinfo {author} {\bibfnamefont{A.~J.}\ \bibnamefont{Buras}}, \bibinfo
  {author} {\bibfnamefont{M.}~\bibnamefont{Jamin}}, \bibinfo {author}
  {\bibfnamefont{M.~E.}\ \bibnamefont{Lautenbacher}},\ and\ \bibinfo {author}
  {\bibfnamefont{P.~H.}\ \bibnamefont{Weisz}},\ }%
  \bibfield{journal}{%
  \Doi{10.1016/0550-3213(93)90397-8}{\bibinfo {journal} {Nucl.Phys.}}\ }%
  \textbf{\bibinfo {volume} {B400}},\ \bibinfo {pages} {37} (\bibinfo {year}
  {1993}),\ \Eprint{http://arxiv.org/abs/hep-ph/9211304}{arXiv:hep-ph/9211304
  [hep-ph]}%
  \bibAnnoteFile{NoStop}{Buras:1992tc}%
\bibitem{Ciuchini:1997bw}%
  \BibitemOpen
  \bibfield{author}{%
  \bibinfo {author} {\bibfnamefont{M.}~\bibnamefont{Ciuchini}}, \bibinfo
  {author} {\bibfnamefont{E.}~\bibnamefont{Franco}}, \bibinfo {author}
  {\bibfnamefont{V.}~\bibnamefont{Lubicz}}, \bibinfo {author}
  {\bibfnamefont{G.}~\bibnamefont{Martinelli}}, \bibinfo {author}
  {\bibfnamefont{I.}~\bibnamefont{Scimemi}}, \emph{et~al.},\ }%
  \bibfield{journal}{%
  \Doi{10.1016/S0550-3213(98)00161-8}{\bibinfo {journal} {Nucl.Phys.}}\ }%
  \textbf{\bibinfo {volume} {B523}},\ \bibinfo {pages} {501} (\bibinfo {year}
  {1998}),\ \Eprint{http://arxiv.org/abs/hep-ph/9711402}{arXiv:hep-ph/9711402
  [hep-ph]}%
  \bibAnnoteFile{NoStop}{Ciuchini:1997bw}%
\bibitem{Buras:1989xd}%
  \BibitemOpen
  \bibfield{author}{%
  \bibinfo {author} {\bibfnamefont{A.~J.}\ \bibnamefont{Buras}}\ and\ \bibinfo
  {author} {\bibfnamefont{P.~H.}\ \bibnamefont{Weisz}},\ }%
  \bibfield{journal}{%
  \Doi{10.1016/0550-3213(90)90223-Z}{\bibinfo {journal} {Nucl.Phys.}}\ }%
  \textbf{\bibinfo {volume} {B333}},\ \bibinfo {pages} {66} (\bibinfo {year}
  {1990})%
  \bibAnnoteFile{NoStop}{Buras:1989xd}%
\bibitem{Buras:1991jm}%
  \BibitemOpen
  \bibfield{author}{%
  \bibinfo {author} {\bibfnamefont{A.~J.}\ \bibnamefont{Buras}}, \bibinfo
  {author} {\bibfnamefont{M.}~\bibnamefont{Jamin}}, \bibinfo {author}
  {\bibfnamefont{M.}~\bibnamefont{Lautenbacher}},\ and\ \bibinfo {author}
  {\bibfnamefont{P.~H.}\ \bibnamefont{Weisz}},\ }%
  \bibfield{journal}{%
  \Doi{10.1016/0550-3213(92)90345-C, 10.1016/0550-3213(92)90345-C}{\bibinfo
  {journal} {Nucl.Phys.}}\ }%
  \textbf{\bibinfo {volume} {B370}},\ \bibinfo {pages} {69} (\bibinfo {year}
  {1992})%
  \bibAnnoteFile{NoStop}{Buras:1991jm}%
\bibitem{Ciuchini:1993vr}%
  \BibitemOpen
  \bibfield{author}{%
  \bibinfo {author} {\bibfnamefont{M.}~\bibnamefont{Ciuchini}}, \bibinfo
  {author} {\bibfnamefont{E.}~\bibnamefont{Franco}}, \bibinfo {author}
  {\bibfnamefont{G.}~\bibnamefont{Martinelli}},\ and\ \bibinfo {author}
  {\bibfnamefont{L.}~\bibnamefont{Reina}},\ }%
  \bibfield{journal}{%
  \Doi{10.1016/0550-3213(94)90118-X}{\bibinfo {journal} {Nucl.Phys.}}\ }%
  \textbf{\bibinfo {volume} {B415}},\ \bibinfo {pages} {403} (\bibinfo {year}
  {1994}),\ \Eprint{http://arxiv.org/abs/hep-ph/9304257}{hep-ph/9304257}%
  \bibAnnoteFile{NoStop}{Ciuchini:1993vr}%
\bibitem{Nakamura:2010zzi}%
  \BibitemOpen
  \bibfield{author}{%
  \bibinfo {author} {\bibfnamefont{K.}~\bibnamefont{Nakamura}} \emph{et~al.}
  (\bibinfo {collaboration} {Particle Data Group}),\ }%
  \bibfield{journal}{%
  \Doi{10.1088/0954-3899/37/7A/075021}{\bibinfo {journal} {J.Phys.G}}\ }%
  \textbf{\bibinfo {volume} {G37}},\ \bibinfo {pages} {075021} (\bibinfo {year}
  {2010})%
  \bibAnnoteFile{NoStop}{Nakamura:2010zzi}%
\bibitem{Martinelli:1994ty}%
  \BibitemOpen
  \bibfield{author}{%
  \bibinfo {author} {\bibfnamefont{G.}~\bibnamefont{Martinelli}}, \bibinfo
  {author} {\bibfnamefont{C.}~\bibnamefont{Pittori}}, \bibinfo {author}
  {\bibfnamefont{C.~T.}\ \bibnamefont{Sachrajda}}, \bibinfo {author}
  {\bibfnamefont{M.}~\bibnamefont{Testa}},\ and\ \bibinfo {author}
  {\bibfnamefont{A.}~\bibnamefont{Vladikas}},\ }%
  \bibfield{journal}{%
  \Doi{10.1016/0550-3213(95)00126-D}{\bibinfo {journal} {Nucl.Phys.}}\ }%
  \textbf{\bibinfo {volume} {B445}},\ \bibinfo {pages} {81} (\bibinfo {year}
  {1995}),\ \Eprint{http://arxiv.org/abs/hep-lat/9411010}{arXiv:hep-lat/9411010
  [hep-lat]}%
  \bibAnnoteFile{NoStop}{Martinelli:1994ty}%
\end{thebibliography}%

\end{document}